\documentclass[10pt,aps,pra,twocolumn,amssymb,amsfonts,floatfix,superscriptaddress,longbibliography,notitlepage,nofootinbib]{revtex4-1}
\usepackage[utf8]{inputenc}
\pdfoutput=1
\usepackage{color}
\usepackage{float}
\usepackage{xcolor}
\usepackage{mathtools,amsmath,amsthm,amsfonts,amssymb}
\usepackage{commath}
\usepackage{physics}
\usepackage{bm}
\usepackage[pdftex]{graphicx}
\usepackage[colorlinks=false]{hyperref}
\usepackage{enumitem}

\begin{document}

\title{Mixed eigenstates in spin-boson systems with one-photon and two-photon interactions}
\author{David Villase{\~n}or}
\email{d.v.pcf.cu@gmail.com}
\author{Marko Robnik}
\email{marko.robnik@guest.um.si}
\affiliation{CAMTP - Center for Applied Mathematics and Theoretical Physics, University of Maribor, Mladinska 3, SI-2000 Maribor, Slovenia, European Union}

\begin{abstract}
Spin-boson systems have attracted increasing attention as accessible experimental platforms and for their potential applications in designing quantum technologies. One characteristic of these systems is the transition from regular to completely chaotic behavior when certain control parameters are varied. However, the characterization of their mixed phase space has not been thoroughly explored. In this work, we investigate the properties of mixed eigenstates in spin-boson systems, comparing one-photon interactions with two-photon interactions. We propose a generalized definition of the phase-space overlap index to identify genuine mixed eigenstates. Our study highlights the fundamental differences that arise when two-photon processes are considered compared to one-photon processes and provides complementary evidence supporting the validity of the principle of uniform semiclassical condensation of quasiprobability functions in spin-boson systems.
\end{abstract}

\maketitle

\section{Introduction}
\label{sec:Introduction}

The study of classical systems with mixed or divided phase space has been crucial in physics, as these systems represent the most typical dynamical motion where regular and chaotic behaviors coexist simultaneously~\cite{LichtenbergBook,OttBook}. In the quantum domain, the origins of describing mixed quantum systems can be traced back to early research that aimed to distinguish between regular and irregular patterns arising in quantum eigenstates~\cite{Percival1973,Berry1977a,Berry1977b,Berry1977JPA}. The insights gained from these studies led to a successful framework for describing mixed quantum systems under the Berry-Robnik picture~\cite{Berry1984} and the subsequent principle of uniform semiclassical condensation (PUSC) of quasiprobability distributions~\cite{Robnik1998,Robnik2007,Robnik2020,Robnik2023}.

The PUSC states that in the semiclassical limit, the eigenstates of a mixed quantum system become strictly regular or chaotic. These eigenstates exhibit spectral statistics that align with the well-established descriptions of regular~\cite{Berry1977a} or chaotic~\cite{Casati1980,Bohigas1984} eigenstates. For mixed quantum systems, the entire spectrum is described by the Berry-Robnik picture~\cite{Berry1984}, whose predictions are supported by substantial numerical evidence~\cite{Prosen1993a,Prosen1993b,Prosen1994a,Prosen1994b,Li1994,Li1995a,Li1995b,Prosen1995,Prosen1998,Prosen1999,Veble1999}. Additionally, the PUSC indicates that in the semiclassical limit, the fraction of mixed eigenstates decays according to a power law~\cite{Robnik1998,Robnik2023}. This decay pattern has been verified in various systems, including the Lemon billiard~\cite{Lozej2022}, the mushroom billiard~\cite{Orel2026b}, the kicked top model~\cite{Wang2023,Yan2024}, the Fermi-Pasta-Ulam-Tsingou model~\cite{Yan2024b}, and the one-photon Dicke model~\cite{Wang2024}.

Spin-boson systems represent an important class of theoretical models that describe the interaction between light and matter~\cite{Scully2008Book}. These systems have gained relevance due to their properties, which are considered valuable for designing and controlling quantum technologies and applications~\cite{Dowling2003,Zeilinger1999,Gisin2007,Swingle2016,Landsman2019,Taylor2016,Degen2017,Fiderer2018,Crawford2021,Montenegro2025,Lloyd1996,Georgescu2014}. In particular, spin-boson Dicke models involving one-photon~\cite{Dicke1954,Kirton2019,Roses2020,Larson2021Book,Villasenor2026ARXIV} and two-photon~\cite{Gerry1989,Ng1999,Emary2002JPA,Larson2021Book} interactions have attracted significant attention. These models illustrate a fundamental cooperative interaction between a set of qubits and an electromagnetic field and have led to fascinating research on superradiant transitions~\cite{Emary2003,Emary2003PRL,Lambert2004,Brandes2005,Brandes2013,Kloc2017,Garbe2017,Chen2018,Ramirez2026}, which hold promise for quantum applications. Moreover, the two-photon processes have spurred further investigation into intriguing phenomena, such as spectral collapse~\cite{Felicetti2015,Duan2016,Rico2020,Lo2021}, photon squeezing~\cite{Gerry1989,Banerjee2022}, dark states~\cite{Peng2017}, and dissipative effects~\cite{Li2022,Li2024,Garbe2020,Shah2025}.

An additional important aspect of spin-boson systems is their relevance to experimental research. Numerous experimental implementations of the one-photon Dicke model have been developed. For a comprehensive review, see Ref.~\cite{Villasenor2026ARXIV} and the references therein. In terms of the two-photon Dicke model, experimental realizations can be traced back to early experiments involving lasers~\cite{Loy1978,Schlemmer1980,Nikolaus1981,Gauthier1992} and masers~\cite{Brune1987,Brune1987PRL,Brune1988}. Contemporary setups now explore the use of Rydberg atoms~\cite{Bertet2002,Zhang2013}, quantum dots~\cite{Stufler2006,DelValle2010}, and quantum batteries~\cite{Crescente2020,Delmonte2021,Wang2024PRA}. Furthermore, interesting setups involving superconducting circuits or trapped ions have provided experimental evidence of spectral collapse~\cite{Felicetti2015,Felicetti2018,Felicetti2019}.

The characterization of classical and quantum chaos has been extensively studied in the one-photon Dicke model~\cite{DeAguiar1992,Bakemeier2013,Bastarrachea2014b,Bastarrachea2015,Chavez2016,Villasenor2026ARXIV}. However, research on chaos in the two-photon Dicke model is currently limited~\cite{Wang2019TPD,Ramirez2025}. While a preliminary investigation into mixed behavior in the one-photon system has been conducted~\cite{Wang2024}, a similar study for the two-photon system is still lacking. Thus, in this work, we examine the properties of the mixed eigenstates in both the one-photon and two-photon Dicke models. We propose a general definition of the phase-space overlap index~\cite{Batistic2013a,Batistic2013b,Batistic2013EPL}, which is currently the primary tool for classifying mixed eigenstates. Our results reveal fundamental differences in mixed eigenstates when considering one-photon versus two-photon processes, further supporting the premise of the PUSC in interacting spin-boson systems.

The remainder of this article is organized as follows. In Sec.~\ref{sec:DickeModel}, we introduce the spin-boson Dicke model, discussing one-photon and two-photon interactions, its general properties, and its classical limit. Section~\ref{sec:Correspondence} explores the connection between the classical and quantum signatures of integrability and chaos in both models, as well as the mixed phase space. In Sec.~\ref{sec:MixedEigenstates}, we present the current methodology to identify mixed eigenstates and introduce the generalized phase-space overlap index. Section~\ref{sec:MixedEigenstatesDicke} shows the analysis of the mixed eigenstates from both one-photon and two-photon models. Finally, in Sec.~\ref{sec:Conclusions}, we summarize our main findings and present our conclusions. Additional details can be found in the Appendixes.

\section{Dicke model with one-photon and two-photon interactions}
\label{sec:DickeModel}

A generalized spin-boson Hamiltonian describing the interaction between photons and a set of $\mathcal{N}$ two-level atoms, setting $\hbar=1$, is given by the Dicke Hamiltonian
\begin{equation}
    \label{eq:DickeHamiltonian}
    \hat{H}_{f} = \omega \hat{a}^{\dagger} \hat{a} + \omega_{0} \hat{J}_{z} + \frac{\gamma}{\mathcal{N}^{f/2}} \left(\hat{a}^{\dagger f} + \hat{a}^{f}\right) \left( \hat{J}_{+} + \hat{J}_{-} \right) ,
\end{equation}
where the operators $\hat{a}^\dagger$ and $\hat{a}$ are the creation and annihilation bosonic operators that satisfy the commutation relation $[\hat{a}, \hat{a}^{\dagger}] = \hat{\mathbb{I}}$. The operators $\hat{J}_{z}$ and $\hat{J}_{\pm}=\hat{J}_{x}\pm i\hat{J}_{y}$ satisfy the commutation relations $[\hat{J}_{z},\hat{J}_{\pm}]=\pm\hat{J}_{\pm}$ and $[\hat{J}_{+},\hat{J}_{-}]=2\hat{J}_{z}$. We define the last operators as collective atomic operators $\hat{J}_{x,y,z} = (1/2)\sum_{k = 1}^{\mathcal{N}}\hat{\sigma}_{x,y,z}^{k}$, where $\hat{\sigma}_{x,y,z}$ are the Pauli matrices.

In Eq.~\eqref{eq:DickeHamiltonian}, the value $f=1$ describes one-photon interactions in the system, while the value $f=2$ describes two-photon interactions. In the first process, one bosonic excitation is created (or annihilated) for each collective atomic excitation. However, in the second process, two bosonic excitations are created (or annihilated) for each collective atomic excitation.

Three main parameters appear in the above Hamiltonian. The radiation bosonic frequency is $\omega$ and the energy splitting between two atomic levels is represented by $\omega_{0}$. The interaction strength between atoms and photons is modulated by the coupling parameter $\gamma$. For the two-photon system $(f=2)$, the spin-boson coupling is limited to a critical value $\gamma_{\text{sc}}=\omega/2$, where the spectral collapse occurs~\cite{Felicetti2015,Duan2016,Rico2020}.

The parity symmetry of the Dicke Hamiltonian is defined by the commutation relation $[\hat{H}_{f},\hat{\Pi}_{f}]=0$, where $\hat{\Pi}_{f} = \text{exp}(i\pi\hat{\Lambda}_{f})$ is the parity operator and $\hat{\Lambda}_{f}=\hat{a}^{\dagger}\hat{a}/f + \hat{J}_{z} + j\hat{\mathbb{I}}$ identifies the generalized number of excitations within the system. For the one-photon system ($f=1$), the parity symmetry corresponds to the simultaneous exchange of the operators $\hat{a} \to - \hat{a}$ and $\hat{J}_x \to -\hat{J}_x$ in the Hamiltonian $\hat{H}_{1}$. The last divides the system into two subspaces identified by the eigenvalues of the parity operator $p = \pm 1$. However, for the two-photon system ($f=2$), the operator exchange is given by $\hat{a} \to i \hat{a}$ and $\hat{J}_x \to -\hat{J}_x$ in the Hamiltonian $\hat{H}_{2}$. For this case, the system is divided into four subspaces specified by the eigenvalues $p=\pm 1,\pm i$.

Both one-photon and two-photon Dicke Hamiltonians possess a conserved quantity. They commute with the squared of the total collective operator, $[\hat{H}_{f},\hat{\mathbf{J}}^{2}]=0$, whose eigenvalues $j(j+1)$ specify different subspaces. In this work, we will use the most studied subspace with value $j=\mathcal{N}/2$.

\subsection{Classical limit of the Dicke model}

Following well-known mean-field approximations with coherent states~\cite{DeAguiar1992,Bakemeier2013,Chavez2016,Villasenor2026ARXIV}, we obtain the classical limit of the Dicke model with one-photon and two-photon interactions. The Glauber and Bloch coherent states of minimum uncertainty~\cite{Arecchi1972,Zhang1990}, $|\alpha\rangle = e^{-|\alpha|^{2}/2}e^{\alpha\hat{a}^{\dagger}}|0\rangle$ and $|\beta\rangle = (1+|\beta|^{2})^{-j}e^{\beta\hat{J}_{+}}|j,-j\rangle$, allow the association of each parameter $(\alpha,\beta)\in\mathbb{C}$ with a pair of canonical position and momentum variables in phase space:
\begin{align}
    \alpha(q,p) & = \sqrt{\frac{j}{2}}(q+ip) , \label{eq:qp1} \\
    \beta(Q,P) & = \frac{Q+iP}{2\Theta(Q,P)} , \label{eq:qp2} \\
    \Theta(Q,P) & = \sqrt{1-\frac{Q^{2}+P^{2}}{4}} ,
\end{align}
where $(q,p)$ are the bosonic classical variables and $(Q,P)$ the atomic classical variables. The factor $\Theta(Q,P)$ adjusts for curvature effects when projecting the Bloch sphere variables onto a plane using a stereographic projection. The above expressions define Glauber-Bloch coherent states in terms of position and momentum classical variables $\mathbf{x}=(q,p;Q,P)$,
\begin{align}
    \label{eq:GlauberState}
    |\mathbf{x}\rangle & \equiv |\alpha\rangle\otimes|\beta\rangle = |q,p\rangle\otimes|Q,P\rangle , \\
    |q,p\rangle & = e^{-(j/4)\left(q^{2}+p^{2}\right)}e^{\left[\sqrt{j/2}(q+ip)\right]\hat{a}^{\dagger}}|0\rangle , \\
    \label{eq:BlochState}
    |Q,P\rangle & = [\Theta(Q,P)]^{2j}e^{\left[(Q+iP)/2\Theta(Q,P)\right]\hat{J}_{+}}|j,-j\rangle ,
\end{align}
where $|0\rangle$ is the photon vacuum and $|j,-j\rangle$ the state with all atoms in the ground state.

The expectation value of the Dicke Hamiltonian in Eq.~\eqref{eq:DickeHamiltonian} using the Glauber-Bloch coherent states produces the classical Hamiltonian~\cite{Villasenor2026ARXIV,Ramirez2025}
\begin{equation}
    \label{eq:ClassicalDickeHamiltonian}
    h_{f}(\mathbf{x}) = \frac{1}{j}\langle\mathbf{x}|\hat{H}_{f}|\mathbf{x}\rangle = \Omega(\mathbf{x}) + \Upsilon(\mathbf{x}) ,
\end{equation}
where 
\begin{equation}
    \Omega(\mathbf{x}) = \frac{\omega}{2}\left(q^{2}+p^{2}\right) + \frac{\omega_{0}}{2}\left(Q^{2}+P^{2}\right)-\omega_{0}
\end{equation}
describes the integrable motion of two decoupled harmonic oscillators and 
\begin{equation}
    \Upsilon(\mathbf{x}) = \frac{2\gamma}{f} \left(q^{f}-(f-1)p^{f}\right) Q \, \Theta(Q,P)
\end{equation}
is a nonlinear coupling term. In Eq.~\eqref{eq:ClassicalDickeHamiltonian}, the scaling by the system size $j$ yields an intensive classical energy $\epsilon=E/j$ and an effective Planck's constant $\hbar_{\text{eff}}=1/j$~\cite{Ribeiro2006}.

\begin{figure*}[ht]
    \centering
    \includegraphics[width=0.9\textwidth]{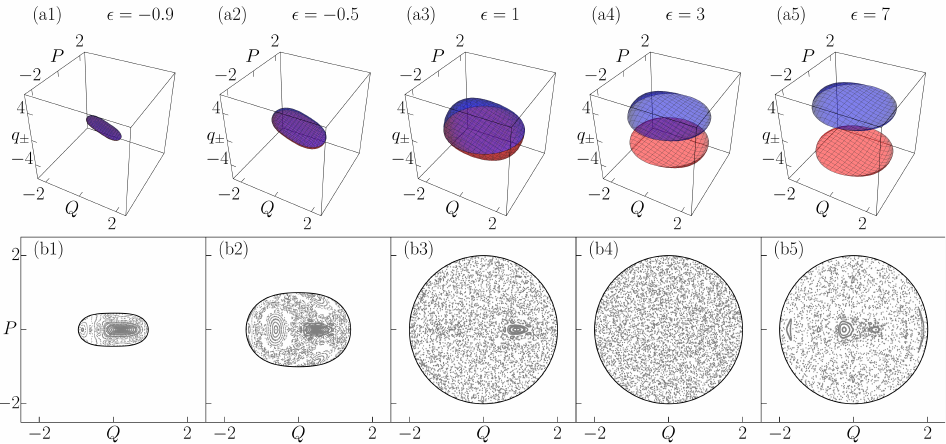}
    \caption{(a1)-(a5) 3D phase-space projections of the classical one-photon Dicke Hamiltonian $h_{1}(\mathbf{x})$ [Eq.~\eqref{eq:ClassicalDickeHamiltonian} with $f=1$]. Each panel represents a different classical energy: (a1) $\epsilon=-0.9$, (a2) $\epsilon=-0.5$, (a3) $\epsilon=1$, (a4) $\epsilon=3$, and (a5) $\epsilon=7$. In (a1)-(a5), the blue (red) surface identifies the positive (negative) $q$ root of the second-degree equation $h_{1}(q,p;Q,P)=\epsilon$. (b1)-(b5) Poincar\'e sections projected on the atomic plane $Q$-$P$ for each value of the classical energies considered in (a1)-(a5). The set of initial conditions is defined by the positive root $q=q_{+}(\epsilon,p=0;Q,P)$ of the second-order equation $h_{1}(q,p;Q,P)=\epsilon$ and the intersection with the plane $p=0$. In (b1)-(b5), the black solid line represents the maximum atomic phase space associated to the classical energy $\epsilon$. System parameters: $\omega=1$, $\omega_0=\omega$, and $\gamma=0.5$.}
    \label{fig:PhaseSpaceOnePhoton}
\end{figure*}

\begin{figure*}[ht]
    \centering
    \includegraphics[width=0.9\textwidth]{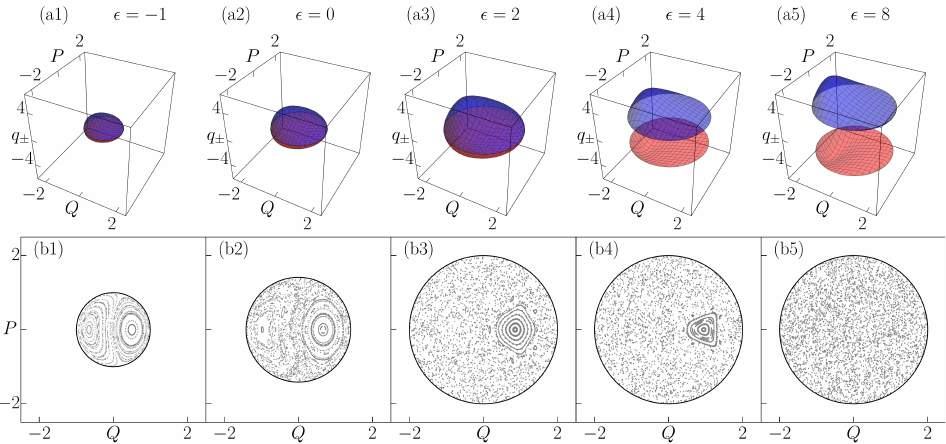}
    \caption{(a1)-(a5) 3D phase-space projections of the classical two-photon Dicke Hamiltonian $h_{2}(\mathbf{x})$ [Eq.~\eqref{eq:ClassicalDickeHamiltonian} with $f=2$]. Each panel represents a different classical energy: (a1) $\epsilon=-1$, (a2) $\epsilon=0$, (a3) $\epsilon=2$, (a4) $\epsilon=4$, and (a5) $\epsilon=8$. In (a1)-(a5), the blue (red) surface identifies the positive (negative) $q$ root of the second-degree equation $h_{2}(q,p;Q,P)=\epsilon$. (b1)-(b5) Poincar\'e sections projected on the atomic plane $Q$-$P$ for each value of the classical energies considered in (a1)-(a5). The set of initial conditions is defined by the positive root $q=q_{+}(\epsilon,p=0;Q,P)$ of the second-order equation $h_{2}(q,p;Q,P)=\epsilon$ and the intersection with the plane $p=0$. In (b1)-(b5), the black solid line represents the maximum atomic phase space associated to the classical energy $\epsilon$. System parameters: $\omega=1$, $\omega_0=2\omega$, and $\gamma=0.3$.}
    \label{fig:PhaseSpaceTwoPhoton}
\end{figure*}

\section{Classical and quantum aspects of chaos}
\label{sec:Correspondence}

Both spin-boson systems that involve one-photon~\cite{Chavez2016, Villasenor2026ARXIV} and two-photon~\cite{Wang2019TPD,Ramirez2025} interactions have exhibited the emergence of classical chaotic dynamics. In this section, we focus on the generalized phase space of both systems to identify the classical transition from integrability to chaotic motion, highlighting areas with mixed dynamics. We also explore how these classical regions are characterized in the quantum realm.

\subsection{Phase space and classical chaos}

We begin by exploring the classical regions where the transition from regular motion to chaotic motion can be observed in both one-photon and two-photon systems. We choose Hamiltonian parameters within a region that allows for a clear visualization of the mixed phase space in each system. The diagram presented in Refs.~\cite{Villasenor2023,Villasenor2026ARXIV} for the one-photon Dicke model helps us select the coupling parameter $\gamma=\gamma_{\text{c}}=\sqrt{\omega_{0}\omega}/2$ at resonance frequencies $\omega_{0}=\omega$. For the two-photon Dicke model, we choose a coupling parameter below the spectral collapse threshold, setting $\gamma = 0.3\omega < \omega/2$ with resonance frequencies at $\omega_{0}=2\omega$~\cite{Ramirez2025}. We use a unity value of the bosonic frequency $\omega=1$ for both systems.

In Fig.~\ref{fig:PhaseSpaceOnePhoton}, we present the phase space of the one-photon Dicke Hamiltonian $h_{1}(\mathbf{x})$ [Eq.~\eqref{eq:ClassicalDickeHamiltonian} with $f=1$]. Figures~\ref{fig:PhaseSpaceOnePhoton}(a1)-\ref{fig:PhaseSpaceOnePhoton}(a5) display 3D projections corresponding to increasing values of the classical energy $\epsilon$. The blue and red surfaces represent the positive and negative $q$ roots, respectively, of the second-degree equation $h_{1}(q,p;Q,P)=\epsilon$. After reaching the classical energy $\epsilon=1$, the two surfaces begin to separate [Figs.~\ref{fig:PhaseSpaceOnePhoton}(a3)-\ref{fig:PhaseSpaceOnePhoton}(a5)], marking the point at which the maximum atomic phase space is attained. 

Figures~\ref{fig:PhaseSpaceOnePhoton}(b1)-\ref{fig:PhaseSpaceOnePhoton}(b5) illustrate Poincar\'e sections projected onto the atomic phase space for each corresponding classical energy $\epsilon$. We evolve a set of initial conditions taking the positive root $q=q_{+}(\epsilon,p=0;Q,P)$ of the second-order equation $h_{1}(q,p;Q,P)=\epsilon$ and the intersection with the plane $p=0$. We observe a transition from integrable motion to chaotic motion, represented by the transformation of ordered patterns into random points. The maximum atomic phase space is achieved at energy $\epsilon=1$ [Fig.~\ref{fig:PhaseSpaceOnePhoton}(b3)]. We can visually identify a zone of mixed dynamics, where chaotic motion starts to emerge smoothly amidst the islands of integrability [Fig.~\ref{fig:PhaseSpaceOnePhoton}(b2)], eventually evolving into a chaotic sea [Fig.~\ref{fig:PhaseSpaceOnePhoton}(b3)]. Interestingly, after entering a region of complete chaotic motion [Fig.~\ref{fig:PhaseSpaceOnePhoton}(b4)], we observe the re-emergence of mixed dynamics as energy increases further [Fig.~\ref{fig:PhaseSpaceOnePhoton}(b5)].

Figure~\ref{fig:PhaseSpaceTwoPhoton} provides a detailed analysis of the two-photon Dicke Hamiltonian $h_{2}(\mathbf{x})$ [Eq.~\eqref{eq:ClassicalDickeHamiltonian} with $f=2$]. In Figs.~\ref{fig:PhaseSpaceTwoPhoton}(a1)-\ref{fig:PhaseSpaceTwoPhoton}(a5), we present 3D projections for increasing values of the classical energy $\epsilon$, where the blue and red surfaces are obtained from the equation $h_{2}(q,p;Q,P)=\epsilon$. For this system, the maximum atomic phase space is attained at an energy of $\epsilon=\omega_{0}$, beyond which the two surfaces separate [Figs.~\ref{fig:PhaseSpaceTwoPhoton}(a3)-\ref{fig:PhaseSpaceTwoPhoton}(a5)]. In Figs.~\ref{fig:PhaseSpaceTwoPhoton}(b1)-\ref{fig:PhaseSpaceTwoPhoton}(b5), we display the Poincar\'e sections corresponding to the previous classical energies. To evolve the initial conditions, we use the positive root $q=q_{+}(\epsilon,p=0;Q,P)$ of the second-order equation $h_{2}(q,p;Q,P)=\epsilon$ and the intersection with the plane $p=0$. We newly observe a transition from integrable dynamics to complete chaotic motion and identify a region of mixed dynamics [Figs.~\ref{fig:PhaseSpaceTwoPhoton}(b2)-\ref{fig:PhaseSpaceTwoPhoton}(b4)].

The percentage of chaotic motion in the previous descriptions can be quantified
with the chaos fraction projected over the atomic plane $Q$-$P$,
\begin{equation}
    \label{eq:ChaosFraction}
    \mu_{c} = \frac{1}{V_{\epsilon}}\int_{\mathcal{M}_{\epsilon}} d\mathbf{x} \, \delta(h_{f}(\mathbf{x})-\epsilon) \, \lambda(Q,P) ,
\end{equation}
where $\lambda$ is a characteristic function that depends on the atomic coordinates $(Q,P)$ and takes two values, $\lambda=1$ for classical regions with positive Lyapunov exponent (chaos) and $\lambda=0$ for regions with zero Lyapunov exponent (integrability). Moreover, $V_{\epsilon} = \int_{\mathcal{M}_{\epsilon}} d\mathbf{x} \, \delta(h_{f}(\mathbf{x})-\epsilon)$ is the volume of the phase space associated to the classical energy $\epsilon$ with subspace $\mathcal{M}_{\epsilon}=\left\{\mathbf{x}=(q,p;Q,P)|h_{f}(\mathbf{x})=\epsilon\right\}$. Thus, $\mu_{c}=1$ defines a fully chaotic classical system, while $\mu_{c}=0$ signifies an integrable classical system.

\begin{figure*}[ht]
    \centering
    \includegraphics[width=0.95\textwidth]{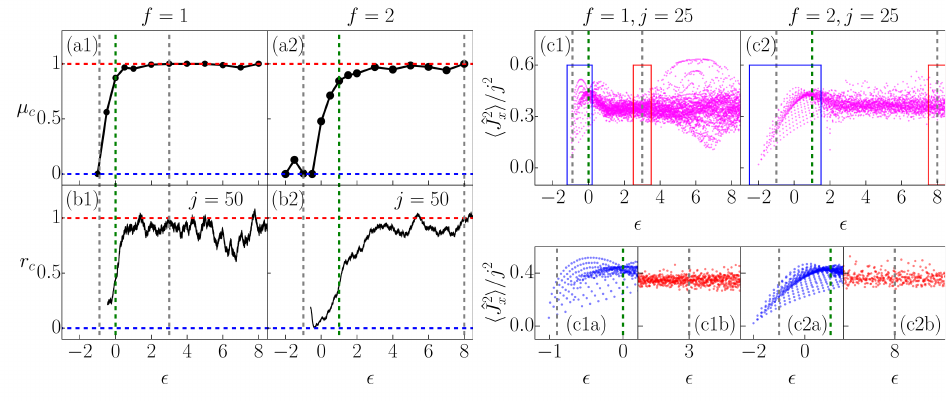}
    \caption{(a1) Chaos fraction $\mu_{c}$ [Eq.~\eqref{eq:ChaosFraction}] for the classical one-photon Dicke Hamiltonian $h_{1}(\mathbf{x})$ [Eq.~\eqref{eq:ClassicalDickeHamiltonian} with $f=1$]. (b1) Normalized spectral ratio $r_{c}$ [Eq.~\eqref{eq:NormSpectralRatio}] for the quantum Hamiltonian $\hat{H}_{1}$ [Eq.~\eqref{eq:DickeHamiltonian} with $f=1$]. (a2)-(b2) Previous quantities for the classical two-photon Dicke Hamiltonian $h_{2}(\mathbf{x})$ [Eq.~\eqref{eq:ClassicalDickeHamiltonian} with $f=2$] and quantum Hamiltonian $\hat{H}_{2}$ [Eq.~\eqref{eq:DickeHamiltonian} with $f=2$]. The spectral ratio was averaged over the corresponding parity sectors of (a1) $\hat{H}_{1}$ ($p=\pm1$) and (a2) $\hat{H}_{2}$ ($p=\pm1,\pm i$). (c1) Peres lattice of the atomic operator $\hat{J}_{x}^{2}$ for the quantum Hamiltonian $\hat{H}_{1}$ with parity $p=1$. (c1a)-(c1b) Enlarged regions (blue and red rectangles) of (c1) to easily visualize the ordered and disordered patterns. (c2), (c2a)-(c2b) Previous lattice for the quantum Hamiltonian $\hat{H}_{2}$ with parity $p=1$. In (b1) and (b2), we use a system size $j=50$, while in (c1) and (c2), we use $j=25$. In (a1) and (a2), the horizontal blue (red) dashed line represents a classical system with integrable (chaotic) dynamics. In (b1) and (b2), the horizontal blue (red) dashed line represents a quantum system with Poisson (GOE) statistics. In (a1)-(c1) [(a2)-(c2)], the vertical gray dashed lines represent regular and chaotic classical energies of $h_{1}(\mathbf{x})$ [Figs.~\ref{fig:PhaseSpaceOnePhoton}(b1) and~\ref{fig:PhaseSpaceOnePhoton}(b4)] and $h_{2}(\mathbf{x})$ [Figs.~\ref{fig:PhaseSpaceTwoPhoton}(b1) and~\ref{fig:PhaseSpaceTwoPhoton}(b5)], respectively. In the same way, the vertical green dashed lines represent two classical energies $\epsilon=0$ and $\epsilon=1$ with mixed dynamics in each system. System parameters ($f=1$): $\omega=1$, $\omega_{0}=\omega$, and $\gamma=0.5$. System parameters ($f=2$): $\omega=1$, $\omega_{0}=2\omega$, and $\gamma=0.3$.}
    \label{fig:PeresLatticeLyapunov}
\end{figure*}

In Fig.~\ref{fig:PeresLatticeLyapunov}(a1), we illustrate the classical transition from integrability to chaotic motion for the one-photon Hamiltonian $h_{1}(\mathbf{x})$. We use the chaos fraction $\mu_{c}$ [Eq.~\eqref{eq:ChaosFraction}] to observe this transition. Figure~\ref{fig:PeresLatticeLyapunov}(a2) presents a similar analysis for the two-photon Hamiltonian $h_{2}(\mathbf{x})$. In both systems, we observe a gradual transition from regular behavior to chaos, with an energy region characterized by mixed dynamics. For the one-photon system, we identify the energy range $\epsilon\in(-1,1]$, while for the two-photon system we see $\epsilon\in[0,4]$, along with fluctuating zones in the range $\epsilon\in[4,8)$. In the following, we will compare these classical transitions with the corresponding quantum transitions in both systems.

\subsection{Signatures of quantum chaos}

After detecting the transition from integrability to chaotic motion in the one-photon and two-photon Dicke models, we now focus on the quantum signatures of this behavior. We employ standard tests based on spectral correlations~\cite{HaakeBook,StockmannBook,WimbergerBook,Gutzwiller1990book}. As established in quantum chaos theory, the presence of chaotic motion in a classical system corresponds to quantum correlations within the quantum system~\cite{Casati1980,Bohigas1984}. These correlations manifest as level repulsion in quantum spectra, which exhibit a distribution of level spacings described by the Gaussian orthogonal ensemble (GOE) distribution from random matrix theory~\cite{Brody1981,Guhr1998,Fyodorov2011,MehtaBook}. The GOE distribution is well approximated by the Wigner-Dyson surmise $P_{\text{GOE}}(s) = (\pi s/2)\text{exp}(-\pi s^{2}/4)$, where $s_k = E_{k+1} - E_k$ is the spacing between two consecutive energy levels. In contrast, classically integrable systems display uncorrelated energy levels in their quantum spectra~\cite{Berry1977a}, which adhere to a spacing distribution described by a Poisson function $P_{\text{P}}(s) = \text{exp}(-s)$.

We simplify our study by considering the spectral ratio~\cite{Oganesyan2007,Atas2013} defined as
\begin{equation}
    \label{eq:SpectralRatio}
    r_{k} = \frac{\min(E_{k+1}-E_{k},E_{k}-E_{k-1})}{\max(E_{k+1}-E_{k},E_{k}-E_{k-1})} ,
\end{equation}
which has proven to be a robust test for quantum chaos due to its ability to avoid the unfolding of quantum spectra~\cite{Guhr1998,HaakeBook,StockmannBook}. The mean value of the spectral ratio $\langle r\rangle$ can effectively distinguish between chaotic and regular quantum systems. For spectra that follow the GOE distribution, the average ratio is given by $\langle r\rangle_{\text{GOE}}=4 - 2\sqrt{3} \approx 0.536$. In contrast, for Poisson-distributed spectra, the average ratio is $\langle r\rangle_{\text{P}}=2\ln2 - 1\approx 0.386$. A normalized spectral ratio can be defined as follows:
\begin{equation}
    \label{eq:NormSpectralRatio}
    r_{c} = \frac{\langle r\rangle - \langle r\rangle_{\text{P}}}{\langle r\rangle_{\text{GOE}} - \langle r\rangle_{\text{P}}} ,
\end{equation}
where $r_{c}=1$ identifies a fully chaotic quantum system, while $r_{c}=0$ represents a regular quantum system.

In Fig.~\ref{fig:PeresLatticeLyapunov}, we illustrate how the integrable and chaotic classical motions identified in the one-photon and two-photon Dicke models translate into the quantum domain. To obtain all sets of eigenvalues and eigenstates from the quantum Hamiltonian $\hat{H}_{f}$ [Eq.~\eqref{eq:DickeHamiltonian}] and perform the spectral analysis, we utilize the efficient basis for the one-photon Dicke Hamiltonian $\hat{H}_{1}$ [Eq.~\eqref{eq:DickeHamiltonian} with $f=1$] and the Fock basis for the two-photon Dicke Hamiltonian $\hat{H}_{2}$ [Eq.~\eqref{eq:DickeHamiltonian} with $f=2$], respectively. For more details on each diagonalization basis, we refer the reader to Appendix~\ref{app:Diagonalization}.

In Fig.~\ref{fig:PeresLatticeLyapunov}(b1), we present the normalized spectral ratio $r_{c}$ [Eq.~\eqref{eq:NormSpectralRatio}] averaged over the two subspaces of the one-photon Hamiltonian $\hat{H}_{1}$. We observe a clear transition from Poisson statistics at low energies to GOE statistics at high energies. The overall quantum transition is well represented by the classical transition shown in Fig.~\ref{fig:PeresLatticeLyapunov}(a1), even with significant quantum fluctuations. The energy interval $\epsilon \in (-1,1]$ in Fig.~\ref{fig:PeresLatticeLyapunov}(b1) exhibits mixed statistics, which correlates with the classical motion illustrated in Figs.~\ref{fig:PhaseSpaceOnePhoton}(b1)-\ref{fig:PhaseSpaceOnePhoton}(b3). Additionally, we find mixed statistics in the energy interval $\epsilon\in[5,8)$. This observation is consistent with the mixed dynamics displayed in Fig.~\ref{fig:PhaseSpaceOnePhoton}(b5), as expected.

Figure~\ref{fig:PeresLatticeLyapunov}(c1) presents the energy distribution of the expectation value of the atomic operator $\hat{J}_{x}^{2}$ with parity $p=1$. This energy distribution is known as the Peres lattice~\cite{Peres1984PRL,Feingold1985,Feingold1986} and serves as a visual test for the loss of integrability. We selected the atomic operator $\hat{J}_{x}^{2}$ for clarity, enabling us to better identify specific patterns within the Peres lattice. We observe a correlation between the ordered patterns in the Peres lattice at low energies and the Poisson statistics indicated by the normalized spectral ratio $r_{c}$. Similarly, the disordered, compressed regions present in the Peres lattices at higher energies align with GOE statistics. For cases of mixed statistics (both Poisson and GOE), we also see a combination of ordered and disordered patterns within the Peres lattice. Figures~\ref{fig:PeresLatticeLyapunov}(c1a)-\ref{fig:PeresLatticeLyapunov}(c1b) show enlarged regions of Fig.~\ref{fig:PeresLatticeLyapunov}(c1) to easily identify the previous patterns. The vertical gray dashed lines shown in Figs.~\ref{fig:PeresLatticeLyapunov}(a1)-\ref{fig:PeresLatticeLyapunov}(c1) and~\ref{fig:PeresLatticeLyapunov}(c1a)-\ref{fig:PeresLatticeLyapunov}(c1b) represent two distinct energies where Poisson and GOE statistics are observed, respectively. These energies correspond to integrable and chaotic motion in the classical limit, as illustrated in Figs.~\ref{fig:PhaseSpaceOnePhoton}(b1) and~\ref{fig:PhaseSpaceOnePhoton}(b4).

Figures~\ref{fig:PeresLatticeLyapunov}(b2)-\ref{fig:PeresLatticeLyapunov}(c2) present the previous quantities for the two-photon Dicke model. In Fig.~\ref{fig:PeresLatticeLyapunov}(b2), we display $r_{c}$ averaged over the four subspaces of the two-photon Hamiltonian $\hat{H}_{2}$. The transition from Poisson statistics at low energies to GOE statistics at high energies is newly detected. The overall agreement between this quantum transition and the classical transition illustrated in Fig.~\ref{fig:PeresLatticeLyapunov}(a2) is satisfactory. Additionally, there exists a substantial energy interval $\epsilon \in [0,7]$ with mixed statistics, which is well described by the classical motion illustrated in Figs.~\ref{fig:PhaseSpaceTwoPhoton}(b2)-\ref{fig:PhaseSpaceTwoPhoton}(b4).

In Figs.~\ref{fig:PeresLatticeLyapunov}(c2) and~\ref{fig:PeresLatticeLyapunov}(c2a)-\ref{fig:PeresLatticeLyapunov}(c2b), we present the Peres lattice of $\hat{J}_{x}^{2}$ with parity $p=1$. We observe a correspondence between $r_{c}$ and the patterns shown by the Peres lattice in the two-photon system, similar to what was identified in the one-photon system. The vertical gray dashed lines in Figs.~\ref{fig:PeresLatticeLyapunov}(a2)-\ref{fig:PeresLatticeLyapunov}(c2) and~\ref{fig:PeresLatticeLyapunov}(c2a)-\ref{fig:PeresLatticeLyapunov}(c2b) indicate two energies characterized by Poisson and GOE statistics, respectively. We can observe the integrable and chaotic motion linked to these classical energies in Figs.~\ref{fig:PhaseSpaceTwoPhoton}(b1) and~\ref{fig:PhaseSpaceTwoPhoton}(b5).

In Appendix~\ref{app:Connection}, we provide additional material about the connection between the average spectral ratio $\langle r\rangle$ and the chaos fraction $\mu_{c}$ for both one-photon and two-photon systems.

\begin{figure*}[ht]
    \centering
    \includegraphics[width=0.95\textwidth]{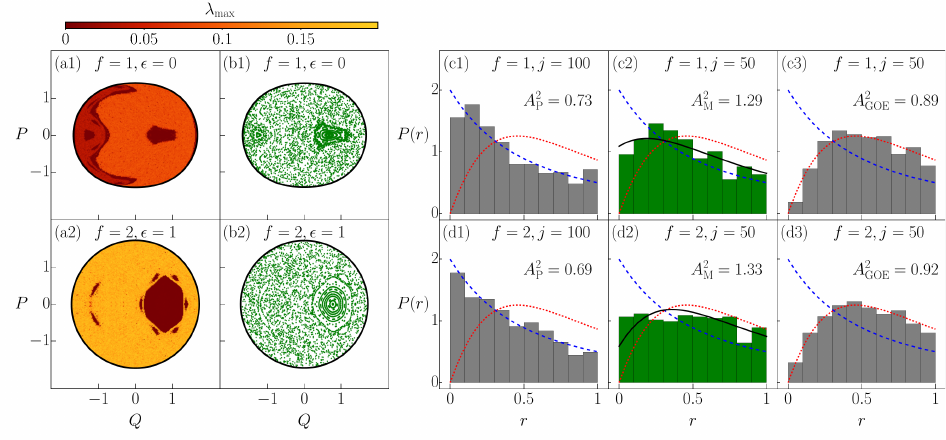}
    \caption{(a1) Maximum Lyapunov exponent and (b1) Poincar\'e section projected on the atomic plane $Q$-$P$ for the classical one-photon Dicke Hamiltonian $h_{1}(\mathbf{x})$ and the classical energy $\epsilon=0$. (a2)-(b2) Previous maps for the classical two-photon Dicke Hamiltonian $h_{2}(\mathbf{x})$ and the classical energy $\epsilon=1$. The top color scale represents the value of the maximum Lyapunov exponent for (a1) and (a2). (c1)-(c3) Statistical distribution of the spectral ratios $r$ [Eq.~\eqref{eq:SpectralRatio}] for the eigenvalues of the quantum Hamiltonian $\hat{H}_{1}$ [Eq.~\eqref{eq:DickeHamiltonian} with $f=1$]. Each panel represents the spectral ratio (around 500-1000 eigenvalues) distribution inside the energy interval $\epsilon_{k}\in[\epsilon-0.1,\epsilon+0.1]$ centered at the classical energies (c1) $\epsilon=-0.9$ (regular), (c2) $\epsilon=0$ (mixed), and (c3) $\epsilon=3$ (chaotic). Figures~\ref{fig:PeresLatticeLyapunov}(a1)-\ref{fig:PeresLatticeLyapunov}(b1) and~\ref{fig:PeresLatticeLyapunov}(c1)-\ref{fig:PeresLatticeLyapunov}(e1) show the previous classical energies as gray (regular), green (mixed), and gray (chaotic) vertical dashed lines, respectively. (d1)-(d3) Statistical distributions for the eigenvalues of the quantum Hamiltonian $\hat{H}_{2}$ [Eq.~\eqref{eq:DickeHamiltonian} with $f=2$]. The energy intervals are centered at the classical energies (d1) $\epsilon=-1$ (regular), (d2) $\epsilon=1$ (mixed), and (d3) $\epsilon=8$ (chaotic). These energies are shown in Figs.~\ref{fig:PeresLatticeLyapunov}(a2)-\ref{fig:PeresLatticeLyapunov}(b2) and~\ref{fig:PeresLatticeLyapunov}(c2)-\ref{fig:PeresLatticeLyapunov}(e2). In (c1)-(d1), we use a system size $j=100$, while in (c2), (c3), (d2), and (d3), we use $j=50$. In (c1)-(d3), the dashed blue (dotted red) line represents a quantum system with Poisson (GOE) statistics [Eqs.~\eqref{eq:SpectralRatioP} and~\eqref{eq:SpectralRatioGOE}]. In (c2)-(d2), the black solid line represents the theoretical mixed-region distribution $P_{\text{M}}(r)$~\cite{Yan2025}. We use the chaos fractions $\mu_{c}=\mu_{c,1}+\mu_{c,2}=0.87$ ($\mu_{c,1}=0.17$, $\mu_{c,2}=0.70$) ($f=1$) and $\mu_{c}=0.84$ ($f=2$) to compute the distributions $P_{\text{M}}(r)$, as described in Ref.~\cite{Yan2025}. In (c1)-(d3), we show the Anderson-Darling parameter $A^{2}$~\cite{Anderson1952}. System parameters ($f=1$): $\omega=1$, $\omega_{0} = \omega$, and $\gamma = 0.5$. System parameters ($f=2$): $\omega=1$, $\omega_{0} = 2\omega$, and $\gamma = 0.3$.}
    \label{fig:SpectralRatio}
\end{figure*}

\subsection{Mixed phase space}

We have found complete correspondence between the spectral statistics and classical motion in the Dicke model with one-photon and two-photon interactions. We observed that the classical transition from integrability to chaotic motion aligns with the quantum transition from Poisson statistics to GOE statistics. Now, we focus on the mixed region, where phase space exhibits mixed classical behavior, and the quantum realm shows a combination of Poisson and GOE statistics, as described by the Berry-Robnik picture~\cite{Berry1984}. We select an appropriate classical energy $\epsilon$ for each system so that the proportion of mixed dynamics remains similar.

The vertical green dashed line in Figs.~\ref{fig:PeresLatticeLyapunov}(a1)-\ref{fig:PeresLatticeLyapunov}(c1) indicates a classical energy characterized by mixed statistics in the one-photon Dicke model, $\epsilon=0$. For convenience, we select a classical energy with a higher fraction of chaotic behavior than regular motion. This choice is made because the presence of multiple regular regions can complicate the analysis by introducing intricate dynamical effects at the boundary between these regions and the chaotic sea~\cite{Tomsovic1994,Doron1995,Backer2008a,Backer2008b,Martinez2021,Vanhaele2022}. Two independent chaotic components emerge in the one-photon system within a very narrow energy interval around $\epsilon\sim(-0.1,0.5)$. The origins of this intriguing classical dynamics deserve further investigation. Thus, for the classical energy $\epsilon=0$, we have two chaotic components with chaos fractions $\mu_{c,1}=0.17$ and $\mu_{c,2}=0.70$, such that the total chaos fraction is $\mu_{c}=\mu_{c,1}+\mu_{c,2}=0.87$.

In Fig.~\ref{fig:SpectralRatio}(a1), we present a map of the maximum Lyapunov exponent projected onto the atomic phase space of the one-photon system. A set of initial conditions was employed to cover this atomic phase space, which we then evolved over a long time period. The color bar at the top indicates the values of the maximum Lyapunov exponent. In this map, we can identify two chaotic components separated by a regular stripe. The first chaotic component has a moonlike shape and a lower value of the maximum Lyapunov exponent compared to the second chaotic component, which occupies the majority of the phase space. Figure~\ref{fig:SpectralRatio}(b1) illustrates the associated Poincar\'e section projected onto the atomic phase space. By comparing both figures, we observe a correlation between regions with a zero Lyapunov exponent and the orderly structures depicted in the Poincar\'e section. In contrast, regions with a positive Lyapunov exponent correspond to a randomlike phase space.

Similarly, in Figs.~\ref{fig:PeresLatticeLyapunov}(a2)-\ref{fig:PeresLatticeLyapunov}(c2), the vertical green dashed line represents an energy with mixed statistics for the two-photon Dicke model, $\epsilon=1$. The chaos fraction for this classical energy is $\mu_{c}=0.84$ and corresponds to a single chaotic component. Figures~\ref{fig:SpectralRatio}(a2)-\ref{fig:SpectralRatio}(b2) illustrate the map of the maximum Lyapunov exponent and the corresponding Poincar\'e section, respectively. The correlation between the values of the maximum Lyapunov exponent and the classical structures shown in the Poincar\'e sections is also confirmed for the two-photon system.

\subsection{Quantum description of mixed regions}

The distributions of the spectral ratio $r$ [Eq.~\eqref{eq:SpectralRatio}], $P(r)$, for Poisson and GOE statistics serve as alternative descriptors for regular and chaotic quantum systems, respectively~\cite{Atas2013},
\begin{align}
    \label{eq:SpectralRatioP}
    P_{\text{P}}(r) & = \frac{2}{(1+r)^{2}} , \\
    \label{eq:SpectralRatioGOE}
    P_{\text{GOE}}(r) & = \frac{27}{4}\frac{r+r^{2}}{(1+r+r^{2})^{5/2}} .
\end{align}
Moreover, a recent study~\cite{Yan2025} has also provided an analytical distribution of the spectral ratio to characterize mixed regions. For a complete description, the functional form, and further details on the derivation of the mixed-region distribution $P_{\text{M}}(r)$, we refer readers to Ref.~\cite{Yan2025}.

In Fig.~\ref{fig:SpectralRatio}, we corroborate the statistics of the spectral ratios for the integrable, mixed, and chaotic regimes in both the one-photon and two-photon Dicke models. In Figs.~\ref{fig:SpectralRatio}(c1)-\ref{fig:SpectralRatio}(c3), we present the statistical distribution of the spectral ratio for the eigenvalues of the one-photon Dicke model. These eigenvalues are contained within an energy interval $\Delta\epsilon$ centered around the classical energies indicated by the vertical dashed lines in Figs.~\ref{fig:PeresLatticeLyapunov}(a1)-\ref{fig:PeresLatticeLyapunov}(c1). The corresponding dynamics of the classical energies confirm the presence of integrable [Fig.~\ref{fig:PhaseSpaceOnePhoton}(b1)], mixed [Fig.~\ref{fig:SpectralRatio}(b1)], and chaotic motion [Fig.~\ref{fig:PhaseSpaceOnePhoton}(b4)], respectively.

We calculate the spectral ratios according to Eq.~\eqref{eq:SpectralRatio} using the eigenvalues that arise from each subspace of the one-photon Hamiltonian $\hat{H}_{1}$. After this calculation, we combine the spectral ratios from each subspace to compute the overall statistics. In Fig.~\ref{fig:SpectralRatio}(c1), we use a larger system size due to the low level density at low energies in spin-boson spectra~\cite{Villasenor2026ARXIV,Ramirez2025}. In Figs.~\ref{fig:SpectralRatio}(c1) and~\ref{fig:SpectralRatio}(c3), we observe that the numerical data align well with the Poisson and GOE statistics associated with regular and chaotic motion. Additionally, Fig.~\ref{fig:SpectralRatio}(c2) illustrates the correspondence between the numerical data and the analytical mixed-region distribution $P_{\text{M}}(r)$, which is depicted as a black solid line. We calculate this distribution using the formula presented in Ref.~\cite{Yan2025} for the case of two chaotic components and one regular component. The implications associated with multiple chaotic components can be studied in detail in Ref.~\cite{Yan2025}. The chaos fractions of the chaotic components are $\mu_{c,1}=0.17$ and $\mu_{c,2}=0.70$, with total chaos fraction $\mu_{c}=\mu_{c,1}+\mu_{c,2}=0.87$. The regularity fraction is obtained by subtraction: $\mu_{r}=1-\mu_{c}$. The agreement between the numerical data and the analytical expressions is further validated using the Anderson-Darling test~\cite{Anderson1952}, where a test parameter is constrained by a threshold $A^{2}\leq2.5$.

A similar analysis for the two-photon Hamiltonian $\hat{H}_{2}$ is shown in Figs.~\ref{fig:SpectralRatio}(d1)-\ref{fig:SpectralRatio}(d3). The energy intervals are centered around the classical energies indicated by the vertical dashed lines in Figs.\ref{fig:PeresLatticeLyapunov}(a2)-\ref{fig:PeresLatticeLyapunov}(c2). The associated dynamics reveal regular [Fig.~\ref{fig:PhaseSpaceTwoPhoton}(b1)], mixed [Fig.~\ref{fig:SpectralRatio}(b2)], and chaotic [Fig.~\ref{fig:PhaseSpaceTwoPhoton}(b5)] motion, respectively. Figures~\ref{fig:SpectralRatio}(d1) and~\ref{fig:SpectralRatio}(d3) confirm the expected Poisson and GOE statistics. In Fig.~\ref{fig:SpectralRatio}(d2), we again observe a good numerical correspondence with the analytical distribution (black solid line) in the mixed region, which was computed for a single chaotic component with chaos fraction $\mu_{c}=0.84$.

\section{Mixed eigenstates}
\label{sec:MixedEigenstates}

\subsection{Husimi function of eigenstates}

We use the quasiprobability Husimi function~\cite{Husimi1940} to study the properties of the eigenstates in the four-dimensional phase space of the Dicke model $\mathcal{M}$ described by the classical variables $\mathbf{x}=(q,p;Q,P)$. We define a generalized Husimi function (unnormalized) with moment $\nu$ for an eigenstate $|E_{k}\rangle$ of the Dicke Hamiltonian [Eq.~\eqref{eq:DickeHamiltonian}] with energy $E_{k}$ as
\begin{equation}
    \mathcal{Q}_{k}^{\nu}(\mathbf{x}) = |\langle E_{k}|\mathbf{x}\rangle|^{2\nu} = |\langle E_{k}|q,p;Q,P\rangle|^{2\nu} ,
\end{equation}
which could be normalized by the constant $A_{\nu}^{k}=\int_{\mathcal{M}} d\mathbf{x}\mathcal{Q}_{k}^{\nu}(\mathbf{x})$. The moment $\nu=1$ provides the standard definition of the Husimi function $\mathcal{Q}_{k}(\mathbf{x})$ with normalization constant $A_{1}^{k}=\int_{\mathcal{M}} d\mathbf{x}\mathcal{Q}_{k}(\mathbf{x})$. Moreover, a Husimi function projected on the atomic plane $Q$-$P$ can be defined as
\begin{equation}
    \label{eq:AtomicHusimi}
    \widetilde{\mathcal{Q}}_{k}^{\nu}(Q,P) = |\langle E_{k}|q=q_{+},p=0;Q,P\rangle|^{2\nu} ,
\end{equation}
which could also be normalized by the constant $B_{\nu}^{k}=\int_{\mathcal{A}} dQdP \, \widetilde{\mathcal{Q}}_{k}^{\nu}(Q,P)$, where $\mathcal{A}=\left\{(Q,P)|Q^{2}+P^{2}\leq4\right\}$ is the atomic phase space. We call $\widetilde{\mathcal{Q}}_{k}^{\nu}(Q,P)$ a Poincar\'e-Husimi function because of the correspondence with the classical Poincar\'e section when defining the bosonic variables as the positive root $q=q_{+}(\epsilon_{k},p=0;Q,P)$ of the second-order equation $h_{f}(q,p;Q,P)=\epsilon_{k}=E_{k}/j$ and the intersection with the plane $p=0$.

\subsection{Phase-space overlap index}

A criterion for selecting mixed eigenstates is based on a measure of the degree of localization of the Husimi function over regular or chaotic regions. This criterion denominated phase-space overlap index was first introduced for the study of mixed eigenstates in billiard systems~\cite{Batistic2013a,Batistic2013b,Batistic2013EPL}. The last test has been shown to be a robust test for selecting mixed eigenstates in different systems~\cite{Lozej2022,Orel2026b,Wang2023,Yan2024,Yan2024b,Wang2024}. However, in this study, we introduce a generalized phase-space overlap index, based on the $\nu$-moment of the Poincar\'e-Husimi function [Eq.~\eqref{eq:AtomicHusimi}]
\begin{equation}
    \label{eq:OverlapIndex}
    M_{\nu} = \frac{1}{B_{\nu}^{k}}\int_{\mathcal{A}} dQdP \, \widetilde{\mathcal{Q}}_{k}^{\nu}(Q,P)\chi(Q,P) ,
\end{equation}
where $B_{\nu}^{k} = \int_{\mathcal{A}} dQdP \, \widetilde{\mathcal{Q}}_{k}^{\nu}(Q,P)$ is the normalization constant and $\chi$ is a classical characteristic function that depends of the atomic coordinates $(Q,P)$. The last function takes two values, $\chi=1$ for classical regions with positive Lyapunov exponent (chaos) and $\chi=-1$ for regions with zero Lyapunov exponent (integrability). The moment $\nu=1$ recovers the standard definition of the phase-space overlap index $M_{1} = (1/B_{1}^{k})\int_{\mathcal{A}} dQdP \, \widetilde{\mathcal{Q}}_{k}(Q,P)\chi(Q,P)$. Thus, $M_{\nu}=1$ corresponds to strictly chaotic eigenstates, $M_{\nu}=-1$ represents strictly regular eigenstates, while mixed eigenstates show overlap-index values inside the interval $M_{\nu}\in(-1,1)$.

The generalized definition of the phase-space overlap index $M_{\nu}$ [Eq.~\eqref{eq:OverlapIndex}] is motivated by the fact that higher moments of the Husimi function can enhance regions where certain eigenstates are localized, while diminishing areas of lower contribution. This enhancement helps to prevent the mistaken overlap of regular and chaotic regions, particularly when the system size is small and the Husimi function is broad. Consequently, our proposed definition can more clearly identify genuine mixed eigenstates. In the following section, we will analyze the behavior of the generalized phase-space overlap index for the eigenstates of the one-photon and two-photon Dicke models, using different moments of the Poincar\'e-Husimi function.

\subsection{Phase-space localization measures}

The localization measures are valuable tools for describing how an eigenstate spreads across a given basis. The generalized R\'enyi-Wehrl entropies have been shown to be effective in characterizing localization in phase space~\cite{Wehrl1978,Gnutzmann2001}. In particular, some studies have focused on the one-photon Dicke model~\cite{Wang2020,Villasenor2021,Pilatowsky2022}. The choice of the subspace in which the localization measure is defined is crucial, as discussed in Ref.~\cite{Villasenor2021}. In this work, we utilize the subspace corresponding to the classical energy $E_{k}$ of a specific eigenstate $|E_{k}\rangle$
\begin{equation}
    \mathcal{M}_{k}=\left\{\mathbf{x}=(q,p;Q,P)|h_{f}(\mathbf{x})=\epsilon_{k}=E_{k}/j\right\} ,
\end{equation}
to define localization measures for the R\'enyi-Wehrl entropies of first and second order~\cite{Villasenor2021}
\begin{align}
    \label{eq:LocMeasure1}
    \mathfrak{L}_{1}(\epsilon_{k}) = & \frac{C_{k}}{V_{k}}\text{exp}\left(-\frac{1}{C_{k}}\int_{\mathcal{M}_{k}}  d\mathbf{s} \,  \mathcal{Q}_{k}(\mathbf{x})\ln\mathcal{Q}_{k}(\mathbf{x})\right) , \\
    \label{eq:LocMeasure2}
    \mathfrak{L}_{2}(\epsilon_{k}) = & \frac{C_{k}^{2}}{V_{k}}\left( \int_{\mathcal{M}_{k}} d\mathbf{s} \, \mathcal{Q}_{k}^{2}(\mathbf{x})\right)^{-1} ,
\end{align}
where $V_{k} = \int_{\mathcal{M}_{k}}d\mathbf{x} \, \delta(h_{f}(\mathbf{x})-\epsilon_{k})= \int_{\mathcal{M}_{k}} d\mathbf{s} \,$ is the phase-space volume associated to the classical energy $\epsilon_{k}$ and we define $d\mathbf{s} = d\mathbf{x} \, \delta(h_{f}(\mathbf{x})-\epsilon_{k})$. The constant $C_{k} = \int_{\mathcal{M}_{k}} d\mathbf{s} \, \mathcal{Q}_{k}(\mathbf{x})$ normalizes the Husimi function inside the subspace $\mathcal{M}_{k}$. The measures $\mathfrak{L}_{1}$ and $\mathfrak{L}_{2}$ are also called the information-entropy localization measure and the inverse participation ratio, respectively, as presented in previous works~\cite{Wang2020,Lozej2022}.

\begin{figure*}[ht]
    \centering
    \includegraphics[width=0.95\textwidth]{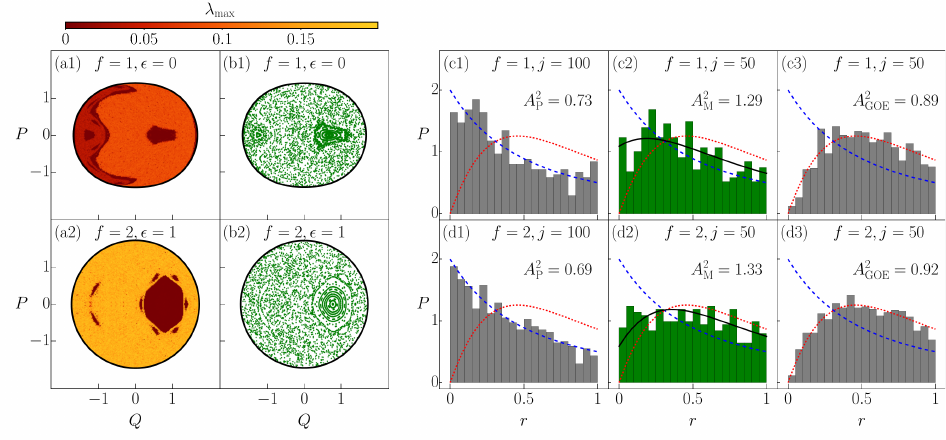}
    \caption{Statistical distribution of the phase-space overlap index $M_{\nu}$ [Eq.~\eqref{eq:OverlapIndex}] for the eigenstates of the one-photon Dicke Hamiltonian $\hat{H}_{1}$ [Eq.~\eqref{eq:DickeHamiltonian} with $f=1$] inside the energy interval $\epsilon \in [-0.05,0.05]$. Each column identifies a different moment of the Husimi function: (a1)-(c1) $\nu=1$, (a2)-(c2) $\nu=2$, (a3)-(c3) $\nu=3$, and (a4)-(c4) $\nu=4$. Each row identifies a different system size: (a1)-(a4) $j\in[20,30]$, (b1)-(b4) $j\in[49,51]$, and (c1)-(c4) $j=100$. (d1)-(d4), (e1)-(e4), and (f1)-(f4) Previous distributions for the eigenstates of the two-photon Dicke Hamiltonian $\hat{H}_{2}$ [Eq.~\eqref{eq:DickeHamiltonian} with $f=2$] inside the energy interval $\epsilon \in [0.95,1.05]$. System parameters ($f=1$): $\omega=1$, $\omega_{0}=\omega$, and $\gamma=0.5$. System parameters ($f=2$): $\omega=1$, $\omega_{0}=2\omega$, and $\gamma=0.3$.}
    \label{fig:OverlapIndex}
\end{figure*}

\begin{figure*}[ht]
    \centering
    \includegraphics[width=0.95\textwidth]{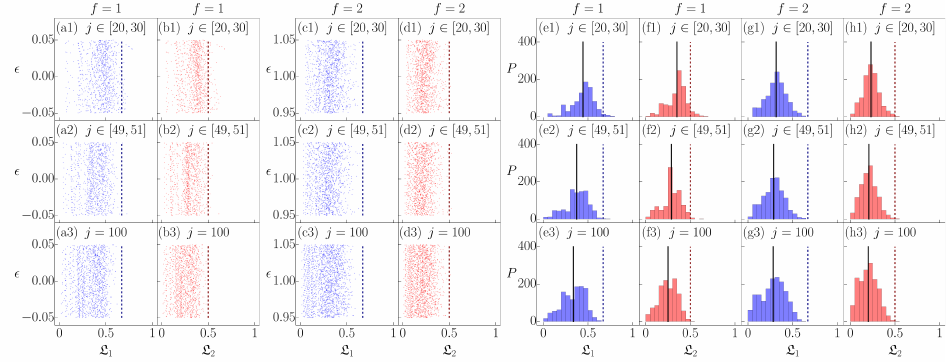}
    \caption{(a1)-(a3) Classical energy $\epsilon$ vs phase-space localization measure $\mathfrak{L}_{1}$ [Eq.~\eqref{eq:LocMeasure1}] for the eigenstates of the one-photon Dicke Hamiltonian $\hat{H}_{1}$ [Eq.~\eqref{eq:DickeHamiltonian} with $f=1$] inside the energy interval $\epsilon \in [-0.05,0.05]$. (b1)-(b3) Classical energy $\epsilon$ vs phase-space localization measure $\mathfrak{L}_{2}$ [Eq.~\eqref{eq:LocMeasure2}] for the previous eigenstates. Each row identifies a different system size: (a1)-(b1) $j\in[20,30]$, (a2)-(b2) $j\in[49,51]$, and (a3)-(b3) $j=100$. (c1)-(c3), (d1)-(d3) Previous observables for the eigenstates of the two-photon Dicke Hamiltonian $\hat{H}_{2}$ [Eq.~\eqref{eq:DickeHamiltonian} with $f=2$] inside the energy interval $\epsilon \in [0.95,1.05]$. (e1)-(e3) Statistical distribution of the phase-space localization measure $\mathfrak{L}_{1}$ for the eigenstates of the one-photon Dicke Hamiltonian $\hat{H}_{1}$ inside the energy interval $\epsilon \in [-0.05,0.05]$. (f1)-(f3) Statistical distribution of the phase-space localization measure $\mathfrak{L}_{2}$ for the previous eigenstates. (g1)-(g3), (h1)-(h3) Previous distributions for the eigenstates of the two-photon Dicke Hamiltonian $\hat{H}_{2}$ inside the energy interval $\epsilon \in [0.95,1.05]$. In (a1)-(a3), (c1)-(c3), (e1)-(e3), and (g1)-(g3), the vertical blue dashed line represents the theoretical value $\mathfrak{L}_{1}^{\max}\approx0.66$ for the maximally delocalized state. In (b1)-(b3), (d1)-(d3), (f1)-(f3), and (h1)-(h3), the vertical red dashed line represents the theoretical value $\mathfrak{L}_{2}^{\max}=1/2$ for the maximally delocalized state. In (e1)-(h3), the vertical black solid line represents the mean value of the localization measures $\mathfrak{L}_{1}$ and $\mathfrak{L}_{2}$ inside the energy interval corresponding to the one-photon or two-photon system. System parameters ($f=1$): $\omega=1$, $\omega_{0} = \omega$, and $\gamma = 0.5$. System parameters ($f=2$): $\omega=1$, $\omega_{0} = 2\omega$, and $\gamma = 0.3$.}
    \label{fig:EnergyLocalization}
\end{figure*}

\begin{figure*}[ht]
    \centering
    \includegraphics[width=0.95\textwidth]{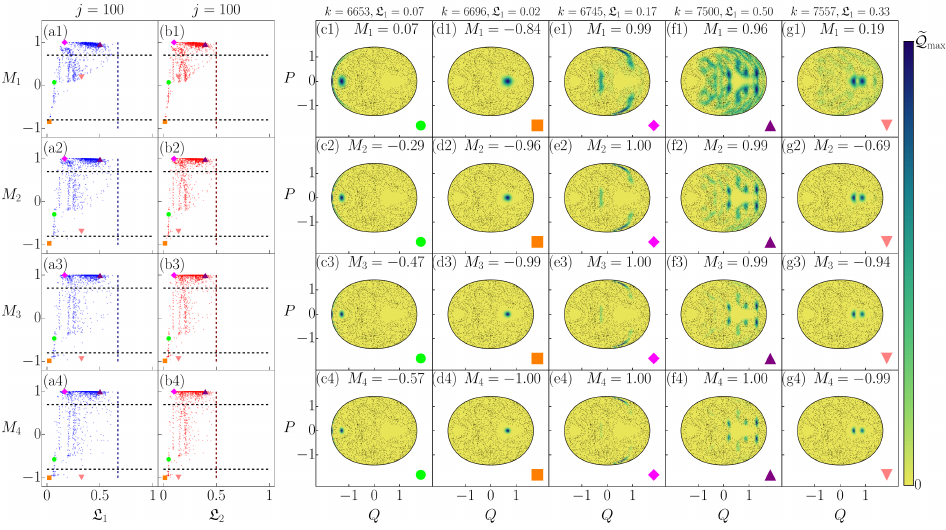}
    \caption{(a1)-(a4) Phase-space overlap index $M_{\nu}$ [Eq.~\eqref{eq:OverlapIndex}] vs phase-space localization measure $\mathfrak{L}_{1}$ [Eq.~\eqref{eq:LocMeasure1}] for the eigenstates of the one-photon Dicke Hamiltonian $\hat{H}_{1}$ [Eq.~\eqref{eq:DickeHamiltonian} with $f=1$] inside the energy interval $\epsilon \in [-0.05,0.05]$. (b1)-(b4) Phase-space overlap index $M_{\nu}$ vs phase-space localization measure $\mathfrak{L}_{2}$ [Eq.~\eqref{eq:LocMeasure2}]. Each row identifies a different moment of the Husimi function: (a1)-(b1) $\nu=1$, (a2)-(b2) $\nu=2$, (a3)-(b3) $\nu=3$, and (a4)-(b4) $\nu=4$. In (a1)-(a4) [(b1)-(b4)], the vertical blue (red) dashed line represents the theoretical value $\mathfrak{L}_{1}^{\max}\approx0.66$ ($\mathfrak{L}_{2}^{\max}=1/2$) for the maximally delocalized state, and the colored dots represent five eigenstates with different properties. In (a1)-(b4), the horizontal black dashed lines represent the interval $\Delta M=M_{+}-M_{-}$. We choose $M_{+}=0.7$ and $M_{-}=-0.8$ for all moments $\nu$. Poincar\'e-Husimi function projected over the atomic phase space $\widetilde{Q}_{k}^{\nu}(Q,P)$ [Eq.~\eqref{eq:AtomicHusimi}] for selected eigenstates: (c1)-(c4) $k=6653$, (d1)-(d4) $k=6696$, (e1)-(e4) $k=6745$, (f1)-(f4) $k=7500$, and (g1)-(g4) $k=7557$. Each row identifies a different moment $\nu$ as in the previous panels. The color scale on the right identifies the value of the Husimi function. In (c1)-(f4), we show the values $M_{\nu}$ and $\mathfrak{L}_{1}$ of the corresponding eigenstate. The small black dots represent the regions of phase space associated with the chaotic motion. System parameters ($f=1$): $\omega=1$, $\omega_{0}=\omega$, $\gamma=0.5$, and $j=100$.}
    \label{fig:HusimiOnePhoton}
\end{figure*}

\begin{figure*}[ht]
    \centering
    \includegraphics[width=0.95\textwidth]{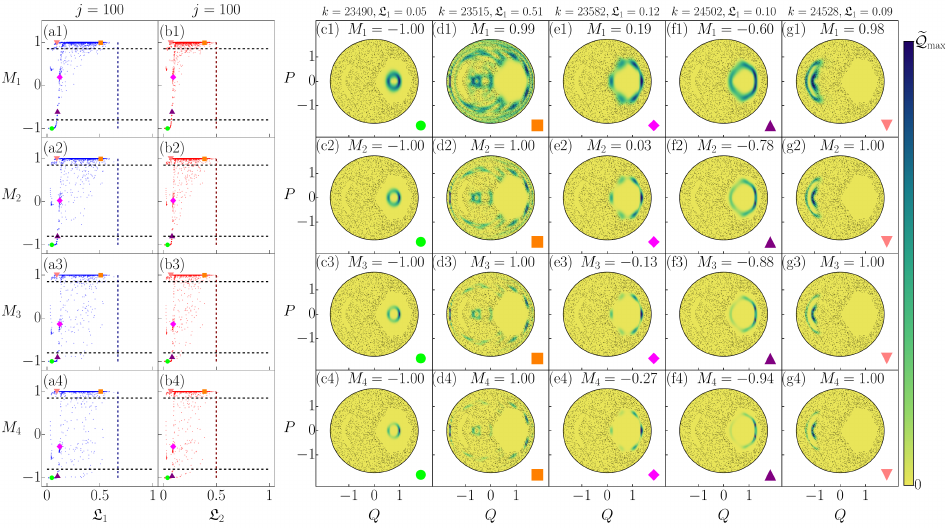}
    \caption{(a1)-(a4) Phase-space overlap index $M_{\nu}$ [Eq.~\eqref{eq:OverlapIndex}] vs phase-space localization measure $\mathfrak{L}_{1}$ [Eq.~\eqref{eq:LocMeasure1}] for the eigenstates of the two-photon Dicke Hamiltonian $\hat{H}_{2}$ [Eq.~\eqref{eq:DickeHamiltonian} with $f=2$] inside the energy interval $\epsilon \in [0.95,1.05]$. (b1)-(b4) Phase-space overlap index $M_{\nu}$ vs phase-space localization measure $\mathfrak{L}_{2}$ [Eq.~\eqref{eq:LocMeasure2}]. Each row identifies a different moment of the Husimi function: (a1)-(b1) $\nu=1$, (a2)-(b2) $\nu=2$, (a3)-(b3) $\nu=3$, and (a4)-(b4) $\nu=4$. In (a1)-(c4) [(b1)-(b4)], the vertical blue (red) dashed line represents the theoretical value $\mathfrak{L}_{1}^{\max}\approx0.66$ ($\mathfrak{L}_{2}^{\max}=1/2$) for the maximally delocalized state, and the colored dots represent five eigenstates with different properties. In (a1)-(b4), the horizontal black dashed lines represent the interval $\Delta M=M_{+}-M_{-}$. We choose $M_{+}=0.85$ and $M_{-}=-0.8$ for all moments $\nu$. Poincar\'e-Husimi function projected over the atomic phase space $\widetilde{Q}_{k}(Q,P)$ [Eq.~\eqref{eq:AtomicHusimi}] for selected eigenstates: (c1)-(c4) $k=23490$, (d1)-(d4) $k=23515$, (e1)-(e4) $k=23582$, (f1)-(f4) $k=24502$, and (g1)-(g4) $k=24528$. Each row identifies a different moment $\nu$ as in the previous panels. The color scale on the right identifies the value of the Husimi function. In (c1)-(f4), we show the values $M_{\nu}$ and $\mathfrak{L}_{1}$ of the corresponding eigenstate. The small black dots represent the regions of phase space associated with the chaotic motion. System parameters ($f=2$): $\omega=1$, $\omega_{0}=2\omega$, $\gamma=0.3$, and $j=100$.}
    \label{fig:HusimiTwoPhoton}
\end{figure*}

\begin{figure*}[ht]
    \centering
    \includegraphics[width=0.95\textwidth]{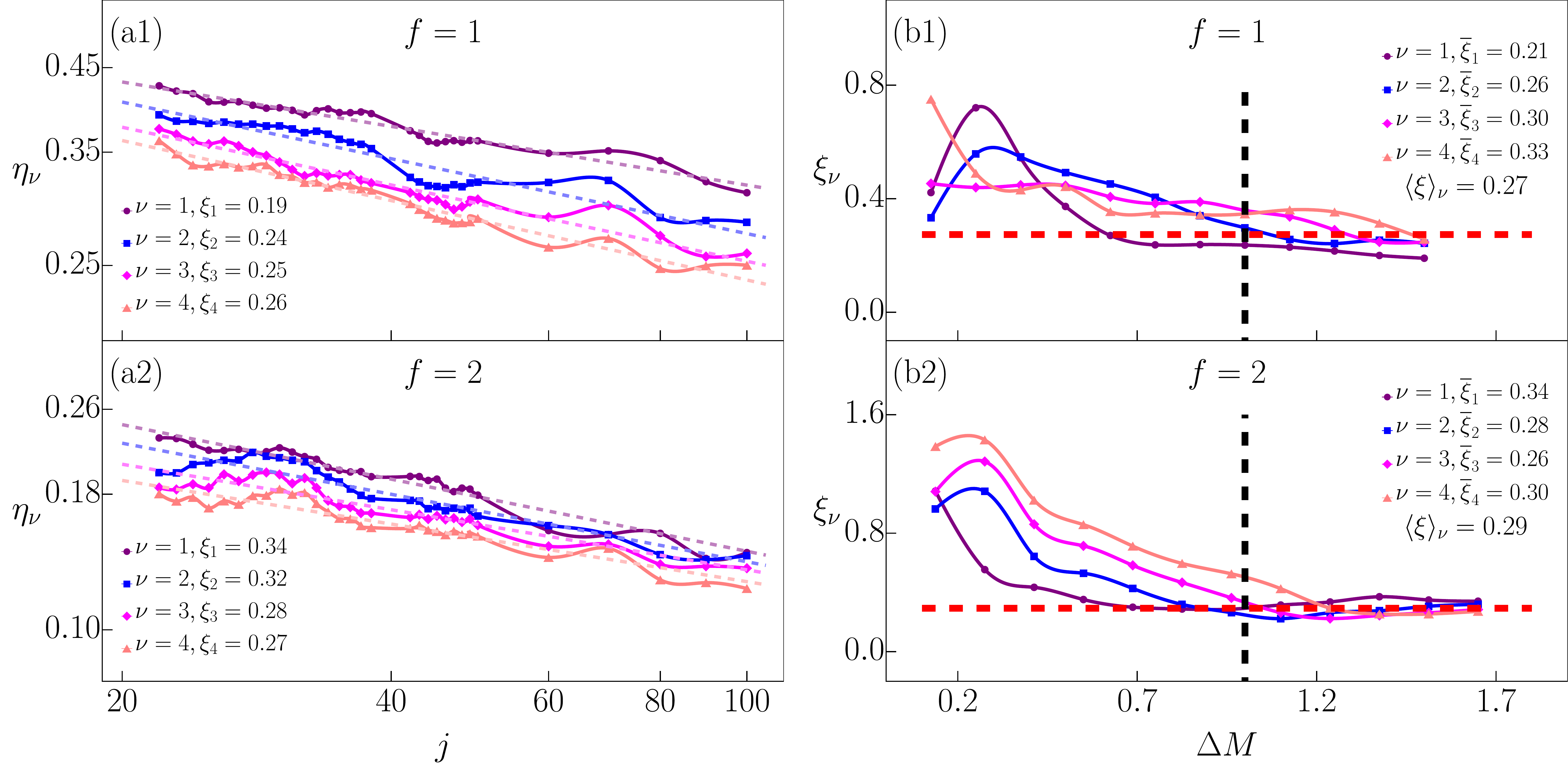}
    \caption{(a1) Power-law decay of the mixed-eigenstate fraction $\eta_{\nu}$ [Eq.~\eqref{eq:EigenstateFraction}] for the eigenstates of the one-photon Dicke Hamiltonian $\hat{H}_{1}$ [Eq.~\eqref{eq:DickeHamiltonian} with $f=1$] inside the energy interval $\epsilon \in [-0.05,0.05]$. (b1) Variation of the decay exponent $\xi_{\nu}$ as a function of the interval $\Delta M = M_{+} - M_{-}$. We fix the upper bound $M_{+}=0.7$ and vary the lower bound inside the interval $M_{-}\in[-0.8,0.5)$. (a2)-(b2) Previous observables for the eigenstates of the two-photon Dicke Hamiltonian $\hat{H}_{2}$ [Eq.~\eqref{eq:DickeHamiltonian} with $f=2$] inside the energy interval $\epsilon \in [0.95,1.05]$. In (b2), we fix the upper bound $M_{+}=0.85$ and vary the lower bound inside the interval $M_{-}\in[-0.8,0.65)$. In (a1) and (a2), the solid lines represent the numerical data, while the dashed lines represent the best linear fit of the function $\ln(\eta_{\nu})=\ln(A)-\xi_{\nu}\ln(j)$. We use a logarithmic scale, such that the slope corresponds to the decay exponent $\xi_{\nu}$ of the function $\eta_{\nu}=Aj^{-\xi_{\nu}}$. In (a1), (a2), (b1), and (b2), each color represents the numerical data computed with a different moment $\nu=1,2,3$, and 4. In (b1) and (b2), the vertical black dashed line signals the large $\Delta M$ region, where the decay exponent $\overline{\xi}_{\nu}$ is averaged for each moment $\nu$. The horizontal red dashed line is the decay exponent $\langle\xi\rangle_{\nu}$ averaged over the moment $\nu$ inside the same region. System parameters ($f=1$): $\omega=1$, $\omega_{0} = \omega$, and $\gamma = 0.5$. System parameters ($f=2$): $\omega=1$, $\omega_{0} = 2\omega$, and $\gamma = 0.3$.}
    \label{fig:PowerLaw}
\end{figure*}

\section{Mixed eigenstates in the Dicke model}
\label{sec:MixedEigenstatesDicke}

In this section, we present the numerical results obtained for the Dicke model with one-photon and two-photon interactions. We investigate the mixed eigenstates that emerge in each model and characterize their fundamental differences.

\subsection{Phase-space overlap index of eigenstates}

We start by examining the phase-space overlap index for the one-photon and two-photon Dicke models. In spin-boson systems, we must consider energy intervals around a classical energy $\epsilon$ to ensure that a sufficient number of eigenvalues (eigenstates) are included for analysis. For the one-photon system, we select the energy interval $\epsilon\in [-0.05,0.05]$ centered at $\epsilon=0$, and for the two-photon system, we choose $\epsilon\in [0.95,1.05]$ centered at $\epsilon=1$. In each case, the classical dynamics remains largely unchanged throughout the entire interval, allowing us to treat it as equivalent to the dynamics at the corresponding energy centers $\epsilon=0$ [Fig.~\ref{fig:SpectralRatio}(b1)] and $\epsilon=1$ [Fig.~\ref{fig:SpectralRatio}(b2)].

As previously discussed, spin-boson spectra exhibit a low level density at low energies, which increases at higher energies~\cite{Villasenor2026ARXIV,Ramirez2025}. One effective method to enhance the level density at low energies is by increasing the system size, as the quantum spectra tend to become continuous in the semiclassical limit. To conduct a systematic analysis, we will begin with smaller system sizes and consider an ensemble average. In our study, we select system sizes of $j\in[20,30]$, $j\in[49,51]$, and $j=100$. For the largest system size $j=100$, the level density is sufficiently high that we do not need to apply an ensemble average.

In Fig.~\ref{fig:OverlapIndex}, we illustrate the statistical distribution of the overlap index $M_{\nu}$ [Eq.~\eqref{eq:OverlapIndex}] for the eigenstates of both one-photon and two-photon systems. For each system, we analyze the eigenstates contained within the respective energy interval, employing the system sizes discussed earlier. We examine four moments of the Husimi function, specifically $\nu=1,2,3,$ and $4$.

In the one-photon system, we observe two main effects related to the increase (from top to bottom) in system size $j$ and (from left to right) in the moment $\nu$. Figures~\ref{fig:OverlapIndex}(a1)-\ref{fig:OverlapIndex}(c1) reveal that at the smallest system size [Fig.~\ref{fig:OverlapIndex}(a1)], there are no regular eigenstates with $M_{1}=-1$. However, as the system size increases [Fig.~\ref{fig:OverlapIndex}(c1)], some eigenstates approach $M_{1}\approx-1$. Furthermore, Figs.~\ref{fig:OverlapIndex}(a1)-\ref{fig:OverlapIndex}(a4) illustrate this effect's dependency on the moment, highlighting eigenstates that exhibit $M_{4}\approx-1$ [Fig.~\ref{fig:OverlapIndex}(a4)]. For the largest system size and moment [Fig.~\ref{fig:OverlapIndex}(c4)], we find that the distinction between purely regular and chaotic eigenstates is sufficiently pronounced, allowing us to classify the remaining eigenstates as genuinely mixed.

A similar study is presented in Figs.~\ref{fig:OverlapIndex}(d1)-\ref{fig:OverlapIndex}(d4),~\ref{fig:OverlapIndex}(e1)-\ref{fig:OverlapIndex}(e4), and~\ref{fig:OverlapIndex}(f1)-\ref{fig:OverlapIndex}(f4) for the two-photon system. However, in this case, increasing the system size or the moment does not appear to have a significant effect. Even at the smallest system size and moment [Fig.~\ref{fig:OverlapIndex}(d1)], we can distinguish between regular and chaotic eigenstates, with $M_{1}=-1$ corresponding to regular states and $M_{1}=1$ corresponding to chaotic states.

The last distinction suggests that the semiclassical limit is more readily achieved in systems with two-photon interactions compared to those with one-photon interactions. This finding prompts us to explore in greater detail the differences between one-photon and two-photon systems.

\subsection{Localization of eigenstates}

We now focus on the localization of eigenstates in the phase space of the one-photon and two-photon Dicke models. We analyze the eigenstates contained within the corresponding energy interval, $\epsilon\in[-0.05,0.05]$ for the one-photon system and $\epsilon\in[0.95,1.05]$ for the two-photon system.

In Fig.~\ref{fig:EnergyLocalization}, we analyze how the localization measures are
distributed across the selected energy interval for both
one-photon and two-photon systems. Figures~\ref{fig:EnergyLocalization}(a1)-\ref{fig:EnergyLocalization}(a3) display the energy distribution of the first-order localization measure $\mathfrak{L}_{1}$ [Eq.~\eqref{eq:LocMeasure1}] in the one-photon system as the system size $j$ increases (from top to bottom). In Fig.~\ref{fig:EnergyLocalization}(a1), which considers the smallest system size, the localization measure $\mathfrak{L}_{1}$ of some eigenstates exceeds the threshold $\mathfrak{L}_{1}^{\max}\approx0.66$ depicted as a dashed line. This value corresponds to a maximum delocalized state (random state)~\cite{Pilatowsky2022}. It becomes evident that the eigenstates with $\mathfrak{L}_{1}>\mathfrak{L}_{1}^{\max}$ diminish in number for the largest system size, as shown in Fig.~\ref{fig:EnergyLocalization}(a3). A similar trend is observed in Figs.~\ref{fig:EnergyLocalization}(b1)-\ref{fig:EnergyLocalization}(b3) for the second-order localization measure $\mathfrak{L}_{2}$ [Eq.~\eqref{eq:LocMeasure2}], where the threshold is $\mathfrak{L}_{2}^{\max}=1/2$.

Additionally, Figs.~\ref{fig:EnergyLocalization}(e1)-\ref{fig:EnergyLocalization}(e3) show the statistical distributions of the localization measure $\mathfrak{L}_{1}$ presented in Figs.~\ref{fig:EnergyLocalization}(a1)-\ref{fig:EnergyLocalization}(a3) for increasing system size. These figures clearly illustrate how the fraction of eigenstates with $\mathfrak{L}_{1}>\mathfrak{L}_{1}^{\max}$ decreases as the system size increases (from top to bottom). The previous effect is also evident when the mean value of the localization measure $\mathfrak{L}_{1}$, represented by a solid line, moves away from the threshold $\mathfrak{L}_{1}^{\max}$ (dashed line) as the system size increases. A similar behavior is also identified in Figs.~\ref{fig:EnergyLocalization}(f1)-\ref{fig:EnergyLocalization}(f3) for the localization measure $\mathfrak{L}_{2}$.

A complementary analysis for the two-photon system is presented in Figs.~\ref{fig:EnergyLocalization}(c1)-\ref{fig:EnergyLocalization}(c3) and~\ref{fig:EnergyLocalization}(d1)-\ref{fig:EnergyLocalization}(d3), which show the energy distribution of the localization measures $\mathfrak{L}_{1}$ and $\mathfrak{L}_{2}$, respectively. Figures~\ref{fig:EnergyLocalization}(g1)-\ref{fig:EnergyLocalization}(g3) and~\ref{fig:EnergyLocalization}(h1)-\ref{fig:EnergyLocalization}(h3) show the corresponding statistical distributions of $\mathfrak{L}_{1}$ and $\mathfrak{L}_{2}$. Here, no significant effect of system size is observed, even for the smallest system size. The last result further emphasizes the feasibility of achieving the semiclassical limit more readily in systems with two-photon interactions.

In Appendix~\ref{app:MIndexLocalization}, we show an additional analysis that explains the connection between the phase-space overlap index $M_{\nu}$ and the localization measures $\mathfrak{L}_{1}$ and $\mathfrak{L}_{2}$ for both one-photon and two-photon systems. The impact of increasing the system size $j$ and the moment $\nu$ is explained in detail, similar to the discussion of the phase-space overlap index in the previous section.

\subsection{Husimi function and mixed eigenstates}

We now turn our attention to the Husimi function of representative eigenstates and the impact of the moment $\nu$. Given the coexistence of regular, chaotic, and mixed eigenstates, we can identify the mixed eigenstates by defining an overlap-index interval
\begin{equation}
    \Delta M = M_{+}-M_{-},
\end{equation}
where $M_{+}<1$ and $M_{-}>-1$ are two boundaries. We can choose symmetric boundaries such that $M_{-}=-M_{+}$, although they are typically asymmetric. Thus, eigenstates with $M_{\nu}\in[-1,M_{-})$ are classified as regular; eigenstates with $M_{\nu}\in(M_{+},1]$ are considered chaotic; and eigenstates with $M_{\nu}\in\Delta M$ are identified as mixed.

In Fig.~\ref{fig:HusimiOnePhoton}, we illustrate the Husimi functions for the one-photon system. For this study, we have chosen the largest system size, $j=100$. Figures~\ref{fig:HusimiOnePhoton}(a1)-\ref{fig:HusimiOnePhoton}(a4) display the lattices $M_{\nu}$ vs $\mathfrak{L}_{1}$ for different moments, $\nu=1,2,3$, and 4. See Appendix~\ref{app:MIndexLocalization} for further details about the impact of system size on the previous lattices. Figures~\ref{fig:HusimiOnePhoton}(b1)-\ref{fig:HusimiOnePhoton}(b4) present the complementary lattices $M_{\nu}$ vs $\mathfrak{L}_{2}$. We have selected five representative eigenstates, each marked with a different color.

We plot the Husimi function for the previous eigenstates projected onto the atomic subspace $\widetilde{Q}_{k}(Q,P)$ [Eq.~\eqref{eq:AtomicHusimi}], examining different moments $\nu$. Figures~\ref{fig:HusimiOnePhoton}(c1)-\ref{fig:HusimiOnePhoton}(c4) illustrate the Husimi functions of an eigenstate characterized by significant localization and an overlap index $M_{1}\in\Delta M$. As we increase the moment $\nu$, the overlap index remains within the interval $M_{\nu}\in\Delta M$, indicating that this eigenstate is a genuine mixed eigenstate. In Figs.~\ref{fig:HusimiOnePhoton}(d1)-\ref{fig:HusimiOnePhoton}(d4), we display the Husimi functions of a regular eigenstate with $M_{1} \in (-1, M_{-})$. For this eigenstate, the overlap index reaches the exact regular value at the largest moment, where $M_{4} = -1$.

In Figs.~\ref{fig:HusimiOnePhoton}(e1)-\ref{fig:HusimiOnePhoton}(e4), we present a scarred eigenstate that exhibits significant localization and an overlap index of approximately $M_{1} \approx 1$. As the moment increases, the region where the eigenstate is localized within the chaotic component expands, resulting in $M_{4} = 1$. In contrast, Figs.~\ref{fig:HusimiOnePhoton}(f1)-\ref{fig:HusimiOnePhoton}(f4) illustrate a chaotic eigenstate characterized by high delocalization and an overlap index also around $M_{1} \approx 1$. With increasing moment, these figures show areas of strong localization within the chaotic component, leading to $M_{4} = 1$.

In Figs.~\ref{fig:HusimiOnePhoton}(g1)-\ref{fig:HusimiOnePhoton}(g4), we present the Husimi functions of an initially mixed eigenstate with $M_{1}\in\Delta M$, which transitions to a regular eigenstate with $M_{4}\approx-1$ as the moment increases. The results indicate that for the lowest moment, $\nu=1$, the Husimi function of this eigenstate overlaps with both chaotic and regular regions. In contrast, at the highest moment, $\nu=4$, the Husimi function is exclusively localized in the regular region. This illustrates the concept of a fictitious mixed eigenstate. We will explore these aspects in the following section.

A similar study is illustrated in Fig.~\ref{fig:HusimiTwoPhoton} for the two-photon system. For the representative eigenstates of this system, we observe behavior analogous to that of the eigenstates in the one-photon system. In Figs.~\ref{fig:HusimiTwoPhoton}(f1)-\ref{fig:HusimiTwoPhoton}(f4), we also identify a fictitious mixed eigenstate with $M_{1}\in\Delta M$, which becomes more regular for the highest moment, $M_{4}\approx-1$.

\subsection{Power-law decay of the mixed-eigenstate fraction}

After characterizing the mixed eigenstates in one-photon and two-photon systems, we investigate the validity of the PUSC. We specifically examine how the relative fraction of mixed eigenstates changes as the system size increases for different moments $\nu$. This generalized mixed-eigenstate fraction is defined by
\begin{equation}
    \label{eq:EigenstateFraction}
    \eta_{\nu} = \frac{N_{\Delta M}}{N_{\Delta \epsilon}} ,
\end{equation}
where $N_{\Delta M}$ represents the number of mixed eigenstates within the interval $\Delta M$, and $N_{\Delta \epsilon}$ denotes the total number of eigenstates within the energy interval $\Delta \epsilon$. According to the PUSC, it has been conjectured that the fraction of mixed eigenstates decays as a power law of the form $\eta_{\nu}\propto j^{-\xi_{\nu}}$ as the system approaches the semiclassical limit. This decay has been confirmed in various systems~\cite{Lozej2022,Orel2026b,Wang2023,Yan2024,Yan2024b,Wang2024} for the moment $\nu=1$, where the decay exponent $\xi_{1}$ can vary under different conditions. Here, we explore the impact of higher moments $\nu>1$ on the generalized exponent $\xi_{\nu}$.

In Fig.~\ref{fig:PowerLaw}, we present an analysis of the PUSC in both one-photon and two-photon systems. Figure~\ref{fig:PowerLaw}(a1) illustrates the power-law decay of the mixed-eigenstate fraction, $\eta_{\nu}$ [Eq.~\eqref{eq:EigenstateFraction}], for the one-photon system. We utilize different moments of the overlap index, $M_{\nu}$, to compute the fraction $\eta_{\nu}$, confirming the power-law decay in all cases. Additionally, we observe that increasing the moment $\nu$ reduces the mixed-eigenstate fraction, consistent with the previous description of the Husimi functions shown in Fig.~\ref{fig:HusimiOnePhoton}. As explained earlier, the increase in the moment eliminates areas of low contribution in the Husimi function, effectively removing spurious overlaps between simultaneous regular and chaotic regions. As a result, the number of mixed eigenstates decreases, leading to a reduced fraction of these states. In Fig.~\ref{fig:PowerLaw}(a2), we present a similar analysis for the two-photon system. Here, we again identify the power-law decay across various moments of the overlap index $M_{\nu}$, mirroring the findings from the one-photon system.

While both one-photon and two-photon systems validate the PUSC, there are some differences between them. Each system exhibits a similar proportion of mixed phase space. The one-photon system has a chaos fraction of $\mu_{c}=0.87$, while the two-photon system has a chaos fraction of $\mu_{c}=0.84$. However, the proportion of mixed eigenstates is higher in the one-photon system [see the vertical scale in Fig.~\ref{fig:PowerLaw}(a1)] compared to the two-photon system [see the vertical scale in Fig.~\ref{fig:PowerLaw}(a2)]. Additionally, when considering small moments, the decay exponent is slightly larger for the two-photon system than for the one-photon system. Specifically, for $\nu=1$: $\xi_{1}=0.19$ ($f=1$) and $\xi_{1}=0.34$ ($f=2$). Conversely, for large moments, both exponents tend to be more similar. For $\nu=4$: $\xi_{4}=0.26$ ($f=1$) and $\xi_{4}=0.27$ ($f=2$).

We now examine how the decay exponents are influenced by variations in the interval $\Delta M = M_{+} - M_{-}$. In Fig.~\ref{fig:PowerLaw}(b1), we present the decay exponent $\xi_{\nu}$ for different moments $\nu$ in the one-photon system. To analyze this variation, we maintain a fixed upper boundary $M_{+}$ while varying the lower boundary $M_{-}$. This approach is taken due to the few quantity of eigenstates within the interval $[M_{-},0]$. See Fig.~\ref{fig:LocalizationOnePhoton} in Appendix~\ref{app:MIndexLocalization} for a complete description.

At small intervals $\Delta M$, we observe that fluctuations are significant, making it difficult to determine a recognizable average decay exponent. In contrast, for larger intervals $\Delta M$, the fluctuations diminish, allowing us to compute an average decay exponent for each moment. We then average the results over the range of large intervals $\Delta M$ to obtain an average exponent $\overline{\xi}_{\nu}$ for each moment $\nu$. Subsequently, we calculate an overall decay exponent defined as $\langle\xi\rangle_{\nu}$ by averaging across different moments.

Additionally, in Fig.~\ref{fig:PowerLaw}(b2), we perform a decay-exponent analysis for the two-photon system. The findings are consistent with those observed in the one-photon system. Both sets of average decay exponents, $\overline{\xi}_{\nu}$ and $\langle\xi\rangle_{\nu}$, in both systems yield values within the interval $\xi_{\nu}\in[0.2,0.35]$. These results align well with decay exponents reported in other studies and systems~\cite{Lozej2022,Orel2026b,Wang2023,Yan2024,Yan2024b,Wang2024}.

\section{Conclusions}
\label{sec:Conclusions}

In this study, we compared the Dicke model with one-photon and two-photon interactions. We selected a mixed phase space, ensuring that each model exhibited a similar proportion of mixed dynamics, and examined the eigenstates of each quantum system. Our focus was on the localization properties of these eigenstates and their corresponding Husimi functions to identify mixed eigenstates within each system. Furthermore, we validated the PUSC of quasiprobability functions in both systems, observing a power-law decay of the mixed-eigenstate fraction as we approached the semiclassical limit. While the PUSC was previously confirmed in the one-photon system~\cite{Wang2024}, this study provides additional support for this finding. Notably, we demonstrate that two-photon processes also adhere to the PUSC. This encourages us to assert that power-law decay of the mixed-eigenstate fraction is a universal characteristic of quantum systems with mixed phase space.

We identified intriguing differences in the Dicke model with one-photon and two-photon interactions. Notably, the one-photon system exhibits a higher fraction of mixed eigenstates compared to the two-photon system, even though both systems display a similar fraction of the classical chaotic component. Although the chaos fractions are similar in both systems, the classical dynamics are not equivalent. For this reason, we believe the previous discrepancy arises from the existence of two chaotic components in the classical limit of the one-photon system, while the two-photon system has only one chaotic component. However, further investigation is needed to understand this difference, along with a more detailed exploration of quantum localization. Another interesting finding is the way the semiclassical limit is achieved in each type of system. Our numerical results suggest that the semiclassical limit is more easily attained in systems with two-photon interactions than in those with one-photon interactions. However, the mechanism by which two-photon interactions facilitate the attainment of the semiclassical limit remains an open question.

A unique aspect of this work is the introduction of the generalized phase-space overlap index $M_{\nu}$, which is based on the $\nu$ moment of the Husimi function. We demonstrated that higher moments, $\nu>1$, also yield a power law that aligns with the standard definition using the moment $\nu=1$. Furthermore, increasing the moment $\nu$ decreases the mixed-eigenstate fraction $\eta_{\nu}$ and slightly modifies the decay exponent $\xi_{\nu}$. The impact of using higher moments of the Husimi function is that it effectively removes regions of minor contribution in phase space. This cleaning process helps eliminate spurious overlaps between regular and chaotic regions caused by broad Husimi functions, allowing for clearer identification of mixed eigenstates. This is particularly advantageous when only small system sizes can be achieved. We also encourage the exploration of the generalized phase-space overlap index $M_{\nu}$ in other systems, and we hope that the analysis presented in this study will be useful for such research.

Several remaining aspects to particularly explore within the context of the two-photon Dicke model include quantum scarring, thermalization, and multifractality, which are already well understood in the one-photon Dicke model~\cite{Pilatowsky2021NatCommun,Villasenor2023,Bastarrachea2024}.

\section*{Acknowledgments}
\label{sec:Acknowledgments}

We acknowledge the help of the supercomputer system HPC Vega - IZUM under project No. S24O02-01. This work was supported by the Slovenian Research and Innovation Agency (ARIS) under Grants No. J1-4387 and No. P1-0306.

\appendix

\section{Numerical solutions and convergence}
\label{app:Diagonalization}

\subsection{Fock basis}

We obtain the solutions of the Dicke Hamiltonian $\hat{H}_{f}$ [Eq.~\eqref{eq:DickeHamiltonian}] through numerical diagonalization. To achieve this, we use the Fock basis, which consists of the tensor product of Fock states $|n\rangle$ (bosonic sector) and angular momentum states $|j,m_{z}\rangle$ (atomic sector),
\begin{equation}
    \label{eq:FockBasis}
    |n;j,m_z\rangle \equiv |n\rangle\otimes|j,m_z\rangle ,
\end{equation}
where $n=0,1,\dots,\infty$ and $m_{z}=-j,-j+1,\ldots,j-1,j$. A maximum photon number $n_{\max}$ should be considered for numerical computations. The Fock basis provides a matrix representation of the Dicke Hamiltonian, with elements defined as follows:
\begin{gather}
    \label{eq:MatrixHamiltonian}
    \langle n';j,m'_{z}|\hat{H}_{f}|n;j,m_{z}\rangle = \left(\omega n + \omega_{0} m_{z}\right)\delta_{n',n}\delta_{m'_{z},m_{z}} \\
    + \frac{\gamma}{\mathcal{N}^{f/2}}\left( A_{n}\delta_{n',n+f} + A_{n-f}\delta_{n',n-f} \right)B_{m'_{z},m_{z}} , \nonumber
\end{gather}
where
\begin{align}
    A_{n} = & \sqrt{(n+1)(n+2)\cdots(n+f)} , \\
    B_{m'_{z},m_{z}} = & C_{m_{z}}^{+}\delta_{m'_{z},m_{z}+1} + C_{m_{z}}^{-}\delta_{m'_{z},m_{z}-1} ,
\end{align}
and $C_{m_{z}}^{\pm} = \sqrt{j(j+1)-m_{z}(m_{z}\pm1)}$.

\subsection{Efficient basis}

The matrix representation in Eq.~\eqref{eq:MatrixHamiltonian} effectively provides the eigenvalues and eigenstates for the Dicke Hamiltonian involving one-photon and two-photon interactions. However, the complexity of the one-photon system necessitates a significant increase in the photon number $n_{\max}$ to explore high-energy regions using the Fock basis [Eq.~\eqref{eq:FockBasis}]. This requirement leads to large matrix dimensions that can exceed available computational resources. Various studies have used efficient bases that enable the exploration of these high-energy regions while minimizing computational demands~\cite{Chen2008,Bastarrachea2014PSa,Bastarrachea2014PSb}.

The efficient basis is written in terms of a displaced bosonic annihilation operator $\hat{A}=\hat{a}+G\hat{J}_{x}$ with $G=2\gamma/(\omega\sqrt{\mathcal{N}})$,
\begin{align}
    |N;j,m_{x}\rangle & = \frac{(\hat{A}^{\dagger})^{N}}{\sqrt{N!}}|\alpha_{m_{x}}\rangle\otimes|j,m_{x}\rangle ,
\end{align}
where $N=0,1,\ldots,\infty$ and $m_{x}=-j,-j+1,\ldots,j-1,j$. A maximum displaced photon number $N_{\max}$ is also selected for numerical computations. In the last expression, $|\alpha_{m_{x}}\rangle=\hat{D}(\alpha_{m_{x}})|0\rangle$ is a coherent state with parameter $\alpha_{m_{x}}=\alpha_{m_{x}}^{\ast}=-Gm_{x}$, and $\hat{D}(\alpha_{m_{x}})=\text{exp}(\alpha_{m_{x}}\hat{a}^{\dagger}-\alpha_{m_{x}}^{\ast}\hat{a})$ is the displacement operator. The states $|j,{m_{x}}\rangle$ are eigenstates of the atomic operator $\hat{J}_x$ with eigenvalue equation $\hat{J}_x|j,{m_{x}}\rangle=m_{x}|j,{m_{x}}\rangle$.

\subsection{Convergence of the solutions}

The truncation values $n_{\max}$ and $N_{\max}$ enable us to construct Hamiltonian matrices with a finite dimension using both the Fock and efficient bases. However, some numerical solutions derived from these matrices can be spurious due to truncation, necessitating the use of a convergence method. In this work, we employ a criterion based on the convergence of the eigenstates, a method that has been widely studied for the one-photon Dicke Hamiltonian \cite{Bastarrachea2014PSa,Bastarrachea2014PSb}, and has recently been examined in the context of the two-photon Dicke model \cite{Ramirez2025}. This convergence criterion proves effective in both the Fock and efficient bases.

The eigenstates of the Dicke Hamiltonian satisfy the eigenvalue equation $\hat{H}_{f}|E_{k}\rangle = E_{k}|E_{k}\rangle$ and can be expanded in a diagonalization basis as
\begin{equation}
    \label{eq:EigenstatesFock}
    |E_{k}\rangle = \sum_{b}c_{b}^{k}|b\rangle,
\end{equation}
where $c_{b}^{k}=\langle b|E_{k}\rangle$. We identify the Fock (efficient) basis when $|b\rangle=|n;j,m_{z}\rangle$ ($|b\rangle=|N;j,m_{x}\rangle$) and define the distribution $p_{n}^{k} = \sum_{m_{z}}|c_{n,m_{z}}^{k}|^{2} \left(P_{N}^{k} = \sum_{m_{x}}|c_{N,m_{x}}^{k}|^{2}\right)$, where $n=0,1,\ldots,n_{\max}$ ($N=0,1,\ldots,N_{\max}$). We evaluate the distribution $p_{n}^{k}$ ($P_{N}^{k}$) at the truncation value $n_{\max}$ ($N_{\max}$), such that $p_{n_{\max}}^{k} \leq \delta$ ($P_{N_{\max}}^{k} \leq \delta$), where $\delta$ is a small tolerance parameter. When the last condition is met, the wavefunction of the corresponding eigenstate remains unchanged within the truncated Hilbert space and is considered well-converged, along with its corresponding eigenvalue. Increasing the truncation value ($n_{\max}$ or $N_{\max}$) does not alter this feature. This criterion is evaluated for each eigenstate, arranged in order of ascending energy.

\begin{figure}[ht]
    \centering
    \includegraphics[width=0.95\columnwidth]{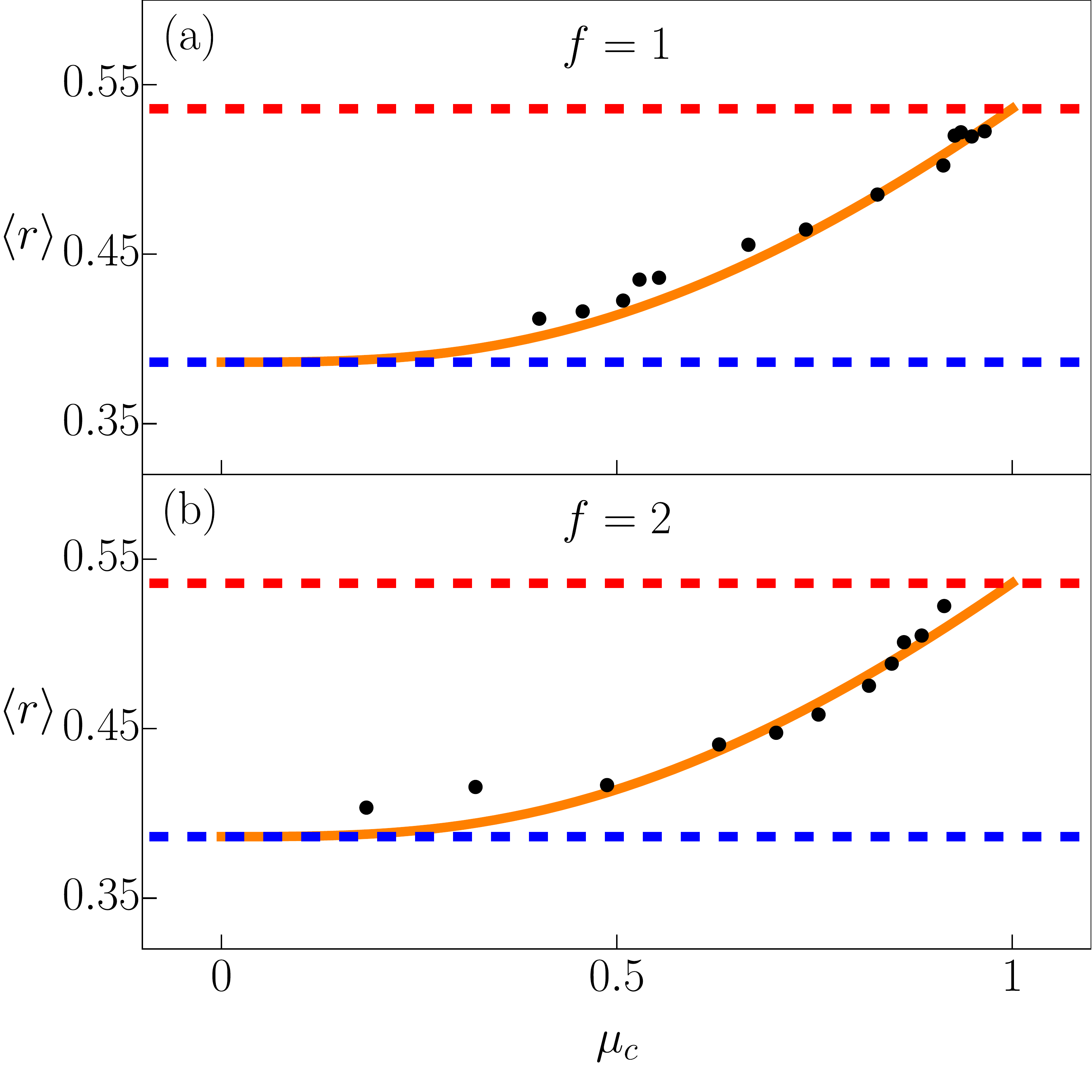}
    \caption{(a) Averaged spectral ratio $\langle r \rangle$ [Eq.~\eqref{eq:SpectralRatio}] as a function of the chaos fraction $\mu_{c}$ [Eq.~\eqref{eq:ChaosFraction}] for the one-photon Dicke model with quantum Hamiltonian $\hat{H}_{1}$ [Eq.~\eqref{eq:DickeHamiltonian} with $f=1$] and classical Hamiltonian $h_{1}(\mathbf{x})$ [Eq.~\eqref{eq:ClassicalDickeHamiltonian} with $f=1$]. (b) Previous quantity for the two-photon Dicke model with quantum Hamiltonian $\hat{H}_{2}$ [Eq.~\eqref{eq:DickeHamiltonian} with $f=2$] and classical Hamiltonian $h_{2}(\mathbf{x})$ [Eq.~\eqref{eq:ClassicalDickeHamiltonian} with $f=2$]. The spectral ratio was averaged over the corresponding parity sectors of (a) $\hat{H}_{1}$ ($p=\pm1$) and (b) $\hat{H}_{2}$ ($p=\pm1,\pm i$), using a system size $j=100$. In (a) and (b), the orange solid line represents the analytical function $\langle r\rangle = \langle r\rangle(\mu_{c})$ developed in Ref.~\cite{Yan2025}, while the black dots are the numerical results. The horizontal blue (red) dashed line represents a quantum system with Poisson (GOE) statistics. System parameters ($f=1$): $\omega=1$, $\omega_{0}=\omega$, and $\gamma=0.5$. System parameters ($f=2$): $\omega=1$, $\omega_{0}=2\omega$, and $\gamma=0.3$.}
    \label{fig:Correspondence}
\end{figure}

\begin{figure*}[ht]
    \centering
    \includegraphics[width=0.95\textwidth]{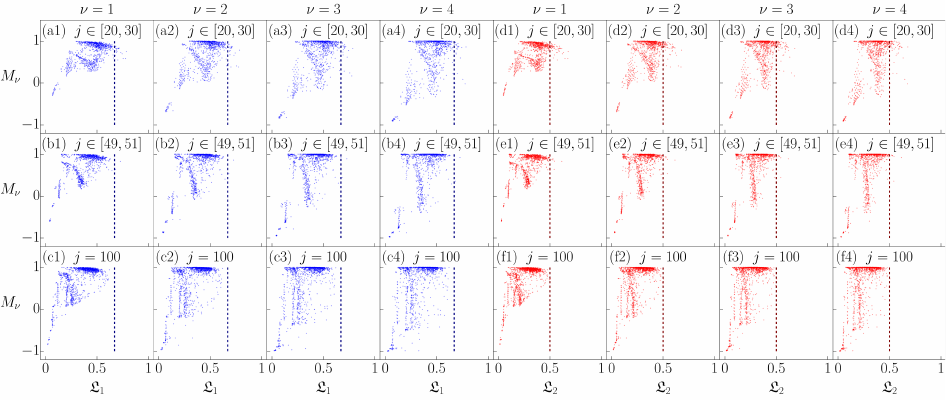}
    \caption{Phase-space overlap index $M_{\nu}$ [Eq.~\eqref{eq:OverlapIndex}] vs phase-space localization measure $\mathfrak{L}_{1}$ [Eq.~\eqref{eq:LocMeasure1}] for the eigenstates of the one-photon Dicke Hamiltonian $\hat{H}_{1}$ [Eq.~\eqref{eq:DickeHamiltonian} with $f=1$] inside the energy interval $\epsilon \in [-0.05,0.05]$. Each column identifies a different moment of the Husimi function: (a1)-(c1) $\nu=1$, (a2)-(c2) $\nu=2$, (a3)-(c3) $\nu=3$, and (a4)-(c4) $\nu=4$. Each row identifies a different system size: (a1)-(a4) $j\in[20,30]$, (b1)-(b4) $j\in[49,51]$, and (c1)-(c4) $j=100$. In (a1)-(c4), the vertical blue dashed line represents the theoretical value $\mathfrak{L}_{1}^{\max}\approx0.66$ for the maximally delocalized state. (d1)-(d4), (e1)-(e4), (f1)-(f4) Phase-space overlap index $M_{\nu}$ vs the phase-space localization measure $\mathfrak{L}_{2}$ [Eq.~\eqref{eq:LocMeasure2}]. In (d1)-(f4), the vertical red dashed line represents the theoretical value $\mathfrak{L}_{2}^{\max}=1/2$ for the maximally delocalized state. System parameters ($f=1$): $\omega=1$, $\omega_{0}=\omega$, and $\gamma=0.5$.}
    \label{fig:LocalizationOnePhoton}
\end{figure*}

\begin{figure*}[ht]
    \centering
    \includegraphics[width=0.95\textwidth]{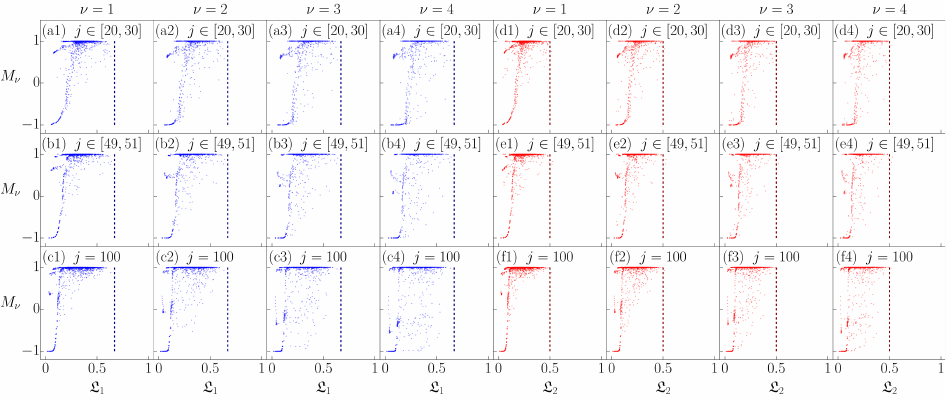}
    \caption{Phase-space overlap index $M_{\nu}$ [Eq.~\eqref{eq:OverlapIndex}] vs phase-space localization measure $\mathfrak{L}_{1}$ [Eq.~\eqref{eq:LocMeasure1}] for the eigenstates of the two-photon Dicke Hamiltonian $\hat{H}_{2}$ [Eq.~\eqref{eq:DickeHamiltonian} with $f=2$] inside the energy interval $\epsilon \in [0.95,1.05]$. Each column identifies a different moment of the Husimi function: (a1)-(c1) $\nu=1$, (a2)-(c2) $\nu=2$, (a3)-(c3) $\nu=3$, and (a4)-(c4) $\nu=4$. Each row identifies a different system size: (a1)-(a4) $j\in[20,30]$, (b1)-(b4) $j\in[49,51]$, and (c1)-(c4) $j=100$. In (a1)-(c4), the vertical blue dashed line represents the theoretical value $\mathfrak{L}_{1}^{\max}\approx0.66$ for the maximally delocalized state. (d1)-(d4), (e1)-(e4), (f1)-(f4) Phase-space overlap index $M_{\nu}$ vs the phase-space localization measure $\mathfrak{L}_{2}$ [Eq.~\eqref{eq:LocMeasure2}]. In (d1)-(f4), the vertical red dashed line represents the theoretical value $\mathfrak{L}_{2}^{\max}=1/2$ for the maximally delocalized state. System parameters ($f=2$): $\omega=1$, $\omega_{0}=2\omega$, and $\gamma=0.3$.}
    \label{fig:LocalizationTwoPhoton}
\end{figure*}

\begin{figure*}[ht]
    \centering
    \includegraphics[width=0.95\textwidth]{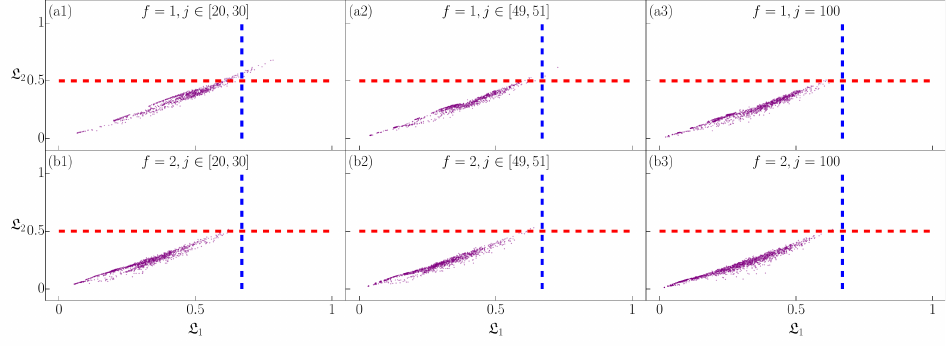}
    \caption{(a1)-(a3) Correlation between the phase-space localization measures $\mathfrak{L}_{1}$ [Eq.~\eqref{eq:LocMeasure1}] and $\mathfrak{L}_{2}$ [Eq.~\eqref{eq:LocMeasure2}] for the eigenstates of the one-photon Dicke Hamiltonian $\hat{H}_{1}$ [Eq.~\eqref{eq:DickeHamiltonian} with $f=1$] inside the energy interval $\epsilon \in [-0.05,0.05]$. (b1)-(b3) Previous correlations for the eigenstates of the two-photon Dicke Hamiltonian $\hat{H}_{2}$ [Eq.~\eqref{eq:DickeHamiltonian} with $f=2$] inside the energy interval $\epsilon \in [0.95,1.05]$. Each column identifies a different system size: (a1)-(b1) $j\in[20,30]$, (a2)-(b2) $j\in[49,51]$, and (a3)-(b3) $j=100$. In (a1)-(b3), the vertical blue (horizontal red) dashed line represents the theoretical value $\mathfrak{L}_{1}^{\max}\approx0.66$ ($\mathfrak{L}_{2}^{\max}=1/2$) for the maximally delocalized state. System parameters ($f=1$): $\omega=1$, $\omega_{0} = \omega$, and $\gamma = 0.5$. System parameters ($f=2$): $\omega=1$, $\omega_{0} = 2\omega$, and $\gamma = 0.3$.}
    \label{fig:Localization}
\end{figure*}

\section{Average spectral ratio and chaos fraction}
\label{app:Connection}

We present in Fig.~\ref{fig:Correspondence} the connection between the average spectral ratio $\langle r\rangle$ [Eq.~\eqref{eq:SpectralRatio}] and the chaos fraction $\mu_{c}$ [Eq.~\eqref{eq:ChaosFraction}] for both one-photon and two-photon Dicke models. In Fig.~\ref{fig:Correspondence}(a), we illustrate the connection for the one-photon system using the corresponding quantum and classical Hamiltonians, $\hat{H}_{1}$ and $h_{1}(\mathbf{x})$ [Eqs.~\eqref{eq:DickeHamiltonian} and~\eqref{eq:ClassicalDickeHamiltonian} with $f=1$]. The orange solid line represents an analytical function that links $\langle r \rangle$ with $\mu_{c}$. This analytical expression was developed in Ref.~\cite{Yan2025} using ensembles of random matrices to simulate the quantum counterpart. We compare the numerical results (black dots) of the one-photon system with the analytical function and find a satisfactory agreement.

The observed deviations can be attributed to the complex structures of quantum spectra in spin-boson systems~\cite{Villasenor2026ARXIV,Ramirez2025}, namely due to some classical stickiness and the corresponding localization of eigenstate Husimi functions, as well as statistical effects. It should be noted that much better agreement between the numerical results and the analytical expression has been reported in one-body systems~\cite{Orel2026b}. Therefore, we are expected to encounter greater deviations in more complex systems, such as spin-boson systems, among others.

In Fig.~\ref{fig:Correspondence}(b), we present the results for the two-photon system, which similarly shows an understandable agreement between the numerical results and the analytical function with certain fluctuations of statistical and dynamical nature.

\section{Phase-space overlap index and phase-space localization measures}
\label{app:MIndexLocalization}

We present the connection between the generalized phase-space overlap index $M_{\nu}$ [Eq.~\eqref{eq:OverlapIndex}] and the localization measures of first and second order, $\mathfrak{L}_{1}$ [Eq.~\eqref{eq:LocMeasure1}] and $\mathfrak{L}_{2}$ [Eq.~\eqref{eq:LocMeasure2}], for both one-photon and two-photon Dicke models.

Figure~\ref{fig:LocalizationOnePhoton} presents a thorough analysis of the localization properties for the eigenstates of the one-photon system within the energy interval $\epsilon\in[-0.05,0.05]$. We calculate the first-order localization measure $\mathfrak{L}_{1}$ and display its distribution alongside the overlap index $M_{\nu}$ for each of the moments $\nu=1,2,3,$ and $4$. We use the same system sizes of $j\in[20, 30]$, $j\in[49, 51]$, and $j = 100$, previously introduced in Sec.~\ref{sec:MixedEigenstatesDicke} for the analysis of the generalized phase-space overlap index $M_{\nu}$. For the smallest system size and moment, shown in Fig.~\ref{fig:LocalizationOnePhoton}(a1), the localization measure $\mathfrak{L}_{1}$ of some eigenstates exceeds the threshold $\mathfrak{L}_{1}^{\max}\approx0.66$ of a maximum delocalized state (random state)~\cite{Pilatowsky2022}. Notably, no eigenstate exhibits $M_{1}=-1$. As we increase (from top to bottom) the system size $j$, illustrated in Fig.~\ref{fig:LocalizationOnePhoton}(c1), there is a reduction in the number of eigenstates with $\mathfrak{L}_{1}>\mathfrak{L}_{1}^{\max}$, while other eigenstates do show $M_{1}=-1$. The increase (from left to right) in the moment $\nu$ plays a significant role in identifying regular states for smaller system sizes, as depicted in Figs.~\ref{fig:LocalizationOnePhoton}(a1)-\ref{fig:LocalizationOnePhoton}(a4). In Fig.~\ref{fig:LocalizationOnePhoton}(a4), we observe that some eigenstates approach values of $M_{4}\approx-1$, while for other eigenstates, $\mathfrak{L}_{1}>\mathfrak{L}_{1}^{\max}$. We confirm a similar pattern for the second-order localization measure $\mathfrak{L}_{2}$ in Figs.~\ref{fig:LocalizationOnePhoton}(d1)-\ref{fig:LocalizationOnePhoton}(d4),~\ref{fig:LocalizationOnePhoton}(e1)-\ref{fig:LocalizationOnePhoton}(e4), and~\ref{fig:LocalizationOnePhoton}(f1)-\ref{fig:LocalizationOnePhoton}(f4). For the last case, the threshold is $\mathfrak{L}_{2}^{\max}=1/2$.

An analogous localization description is presented in Fig.~\ref{fig:LocalizationTwoPhoton} for the eigenstates of the two-photon Dicke model within the energy interval $\epsilon\in[0.95,1.05]$. In this system, we observe that even for small system sizes and moments [Fig.~\eqref{fig:LocalizationTwoPhoton}(a1)], there are no eigenstates with $\mathfrak{L}_{1}>\mathfrak{L}_{1}^{\max}$. In contrast, regular eigenstates exhibit $M_{1}=-1$. Therefore, increasing either the system size [Fig.~\eqref{fig:LocalizationTwoPhoton}(c1)] or the moment [Fig.~\eqref{fig:LocalizationTwoPhoton}(a4)] does not significantly improve the situation in the two-photon system, unlike the one-photon system. A similar pattern for the second-order localization measure $\mathfrak{L}_{2}$ is shown in Figs.~\ref{fig:LocalizationTwoPhoton}(d1)-\ref{fig:LocalizationTwoPhoton}(d4),~\ref{fig:LocalizationTwoPhoton}(e1)-\ref{fig:LocalizationTwoPhoton}(e4), and~\ref{fig:LocalizationTwoPhoton}(f1)-\ref{fig:LocalizationTwoPhoton}(f4). These results support the notion that the semiclassical limit is more easily achieved in systems involving two-photon interactions.

Previous studies have shown that the localization measures of first and second order, $\mathfrak{L}_{1}$ and $\mathfrak{L}_{2}$, are approximately linearly correlated~\cite{Lozej2022,Orel2026b}. In this way, the order of the measure does not affect the overall interpretations of quantum localization. In Fig.~\ref{fig:Localization}, we illustrate the correlation between the two localization measures, $\mathfrak{L}_{1}$ and $\mathfrak{L}_{2}$, for both one-photon and two-photon systems. Figures~\ref{fig:Localization}(a1)-\ref{fig:Localization}(a3) depict the correlation in the one-photon system as the system size increases (from left to right), while Figs.~\ref{fig:Localization}(b1)-\ref{fig:Localization}(b3) show the correlation in the two-photon system. In both systems, we confirm a linear-like correlation between the localization measures. However, in Fig.~\ref{fig:Localization}(a1) for the one-photon system, we detect the effects of finite system size, indicated by values where $\mathfrak{L}_{1}>\mathfrak{L}_{1}^{\max}$ and $\mathfrak{L}_{2}>\mathfrak{L}_{2}^{\max}$. These effects diminish as the system size increases, as shown in Figs.~\ref{fig:Localization}(a2) and~\ref{fig:Localization}(a3). In contrast, this effect is not present in the two-photon system, as evidenced by Figs.~\ref{fig:Localization}(b1)-\ref{fig:Localization}(b3).

\clearpage
\newpage

\bibliography{main}

\end{document}